\documentclass[final,a4paper,3p]{elsarticle}

\journal{arxiv.org}

\usepackage[lined,ruled,linesnumbered,commentsnumbered]{algorithm2e}
\DontPrintSemicolon
\usepackage{array}                   % customize tables and arrays
\usepackage{amsmath,amssymb,amsfonts,amsthm}
\allowdisplaybreaks
\usepackage{dsfont,bm,bbm} 
\usepackage{relsize}
\usepackage{mathrsfs}
\usepackage{booktabs}
\usepackage{mathtools}
\usepackage[german,english]{babel}   % language to use in the document
\usepackage{caption}                 
\usepackage[utf8]{inputenc}          % special characters and encoding
\usepackage[T1]{fontenc}
\usepackage{float}
\usepackage{etoolbox}
\usepackage{graphicx}                % graphics
\usepackage{lmodern}                 % some fonts
\usepackage{microtype}    % controls spaces between words and paragraphs
\usepackage{lineno}
\usepackage{multirow}
\usepackage{calligra}
\usepackage{nicefrac}
\usepackage[usenames,dvipsnames]{xcolor}
\usepackage{tikz}
\usepackage{tikz-network}
\usepackage[caption=false]{subfig}   % get rid of warning 
\usepackage[deletedmarkup=xout,authormarkup=none]{changes}
\usepackage{url} 
\usepackage[colorlinks]{hyperref}
\usepackage{cleveref}
\crefformat{equation}{(#2#1#3)}
\AtBeginDocument{%
  \hypersetup{
    citecolor=Black,
    linkcolor=Black,   
    urlcolor=Violet,
	pdfauthor={Uribe et al.}}}

\ifpdf
\hypersetup{
	pdftitle={sparsity promoting priors},
	pdfauthor={Uribe et al.}
}
\fi

\makeatletter
\renewcommand{\@algocf@capt@plain}{above}% formerly {bottom}
\makeatother

\makeatletter
\def\url@leostyle{%
\@ifundefined{selectfont}{\def\UrlFont{\sf}}{\def\UrlFont{\small\ttfamily}}}
\makeatother
\newcommand*\patchAmsMathEnvironmentForLineno[1]{%
  \expandafter\let\csname old#1\expandafter\endcsname\csname #1\endcsname
  \expandafter\let\csname oldend#1\expandafter\endcsname\csname end#1\endcsname
  \renewenvironment{#1}%
     {\linenomath\csname old#1\endcsname}%
     {\csname oldend#1\endcsname\endlinenomath}}% 
\newcommand*\patchBothAmsMathEnvironmentsForLineno[1]{%
  \patchAmsMathEnvironmentForLineno{#1}%
  \patchAmsMathEnvironmentForLineno{#1*}}%
\AtBeginDocument{%
\patchBothAmsMathEnvironmentsForLineno{equation}%
\patchBothAmsMathEnvironmentsForLineno{align}%
\patchBothAmsMathEnvironmentsForLineno{flalign}%
\patchBothAmsMathEnvironmentsForLineno{alignat}%
\patchBothAmsMathEnvironmentsForLineno{gather}%
\patchBothAmsMathEnvironmentsForLineno{multline}%
}

\newcommand{\given}{\;\ifnum\currentgrouptype=16 \middle\fi|\;}
\newcommand{\suchthat}{\;\ifnum\currentgrouptype=16 \middle\fi|\;}
\newcommand{\norm}[1]{\left\Vert#1\right\Vert}

\newcommand{\dd}{\mathrm{d}}
\newcommand{\tran}{\mathsf{T}}

\newcommand{\ve}[1]{\boldsymbol{#1}}
\newcommand{\matr}[1]{\boldsymbol{#1}}
\newcommand{\scaleT}{\tau}
\newcommand{\stT}{$t$ }

\newcommand{\invg}{\text{IG}}
\newcommand{\gam}{\text{Ga}}
\newcommand{\gamtr}{\text{Ga}_{\nu>1}}
\DeclareMathOperator{\diag}{diag}
\newtheorem{theorem}{Theorem}[section]
\newtheorem{remark}[theorem]{Remark}
\newtheorem{definition}{Definition}

\graphicspath{{figures/}}

\begin{document}
%Scale mixture representations for a class of Markov random field priors: application to edge-preserving Bayesian inversion
% Heavy-tailed Markov random field priors for edge-preserving Bayesian inversion: a scale mixture representation approach
\begin{frontmatter}	
\title{Bayesian inversion with Student's \stT priors based on Gaussian scale mixtures}

\author[LUT]{Angelina Senchukova\corref{cor1}}
\author[LUT]{Felipe Uribe} 
\author[LUT]{Lassi Roininen}

\cortext[cor1]{Corresponding author.} % Tel.: +49 89 289 23017.}

\address[LUT]{School of Engineering Sciences, Lappeenranta--Lahti University of Technology. Yliopistonkatu 34, 53850 Lappeenranta, Finland.}

\begin{abstract}
Many inverse problems focus on recovering a quantity of interest that is a priori known to exhibit either discontinuous or smooth behavior. Within the Bayesian approach to inverse problems, such structural information can be encoded using Markov random field priors. We propose a class of priors that combine Markov random field structure with Student's \stT distribution. This approach offers flexibility in modeling diverse structural behaviors depending on available data. Flexibility is achieved by including the degrees of freedom parameter of Student's \stT distribution in the formulation of the Bayesian inverse problem. To facilitate posterior computations, we employ Gaussian scale mixture representation for the Student's \stT Markov random field prior, which allows expressing the prior as a conditionally Gaussian distribution depending on auxiliary hyperparameters. Adopting this representation, we can derive most of the posterior conditional distributions in a closed form and utilize the Gibbs sampler to explore the posterior. We illustrate the method with two numerical examples: signal deconvolution and image deblurring.
\end{abstract}

\begin{keyword}
Bayesian inverse problems, Bayesian hierarchical modeling, Student's \stT distribution, Markov random fields, Gaussian scale mixture, Gibbs sampler.
\vspace*{5pt}

\noindent \emph{AMS}: 62F15, 65C05, 65R32, 65F22.
\end{keyword}
\end{frontmatter}

%===MAIN TEXT==================================================
\section{Introduction}
Inverse problems aim at recovering a quantity of interest from indirect observations based on a (forward) mathematical model \cite{tarantola_2004}. Depending on the structural features of the unknown to be recovered, inverse problems can be divided into different types. Some problems seek reconstruction of the unknown with discontinuous structures exhibiting large jumps, while others aim at recovering unknown functions that are almost surely continuous. Examples of inverse problems requiring preservation of the edges or sharp features in the solution occur in imaging science, such as X-ray computed tomography, image deblurring, segmentation, and denoising \cite{zhang_2022,springer_2023,hansen_2010}. Conversely, the emphasis in many inverse problems often lies in recovering smooth features, typically when determining coefficients of partial differential equations \cite{marzouk_and_najm_2009,cui_et_al_2014,uribe_et_al_2021}.

One of the characteristics of inverse problems is their instability or ill-posed nature, where even minor perturbations in observations can lead to nonexistence or lack of uniqueness in the solution. To obtain a stable solution, deterministic regularization techniques such as Tikhonov regularization are often used (see, e.g., \cite{hansen_2010}). An alternative approach reformulates inverse problems as problems of statistical inference \cite{kaipio_and_somersalo_2005}. In particular, in Bayesian statistics, the unknown quantity, observations, and noise are treated as random variables, and the solution to the inverse problem is formulated in the form of the posterior distribution.

Bayesian framework naturally allows incorporating prior knowledge about the structure of the unknown, which also acts as a regularizer, in the form of the prior distribution. \emph{Markov random field} (MRF) priors are a useful tool in constructing structural priors (see, e.g., \cite{bardsley_2019}). Such models can be integrated with Student's \stT distribution, offering a flexible prior capable of capturing diverse solution behaviors. Student's \stT distribution can be tuned through its degrees of freedom parameter. With lower degrees of freedom, Student's \stT distribution exhibits heavy-tailed characteristics (sharpness). Conversely, as the degrees of freedom parameter approaches infinity, the distribution converges towards a Gaussian distribution (smoothness).

While providing enhanced flexibility, the combination of MRF priors with Student's \stT distributions can lead to complex posterior distributions that are hard to explore. To alleviate some of these difficulties, we utilize the \emph{Gaussian scale mixture} (GSM) representation of Student's \stT distribution \cite{andrews_and_mallows_1974}. GSMs allow for the representation of a random variable in terms of a conditionally Gaussian variable and a mixing random variable, offering several advantages, particularly in sampling procedures to explore the posterior. By employing the GSM for Student's \stT distribution, we can express it in terms of conditionally Gaussian and inverse-gamma distributions. Consequently, many conditional terms of the resulting hierarchical posterior simplify into products of Gaussian and inverse-gamma densities, enabling closed-form expressions through conjugacy. This construction motivates the adoption of the Gibbs sampler for posterior sampling \cite{geman_and_geman_1984}.

One notable challenge in our approach arises from the inference of the degrees of freedom parameter, as obtaining the conditional posterior for this parameter in closed form is unfeasible. To address this, we employ the classic Metropolis algorithm to sample the conditional within the Gibbs structure, thereby constructing a hybrid Gibbs sampler \cite[p.389]{robert_and_casella_2004}, commonly referred to as Metropolis-within-Gibbs \cite{muller_1991}. Furthermore, the statistical modeling of the degrees of freedom parameter of  Student's \stT distribution is a challenging task. Oftentimes this parameter is not clearly determined by the data. Finding tractable informative prior distributions for such a parameter is necessary. Popular prior choices existing in statistics are exponential prior \cite{geweke_1993, fernandez_and_steel_1998}, Jeffreys prior \cite{fonseca_2008}, gamma prior \cite{juarez_2010}, penalized complexity prior \cite{simpson_et_al_2017}, reference prior \cite{He_2021}, and log-normal prior \cite{lee_2022}. In this work, we also computationally compare some of the previous priors, aiming to offer recommendations tailored to linear inverse problem scenarios.

\paragraph{Related work} 
In linear Bayesian inverse problems, MRF priors are commonly employed to induce structural characteristics in the solution. For instance, Gaussian MRF priors are utilized to introduce smoothing structure \cite{bardsley_2013,roininen_2014}, while piecewise constant structure are modeled using Laplace MRF \cite{bardsley_2012}, Cauchy and alpha-stable MRF \cite{suuronen_et_al_2022,senchukova_2023,suuronen_2023}, and horseshoe MRF \cite{uribe_et_al_2023}. Additionally, hierarchical prior models, as described in \cite{calvetti_and_somersalo_2007, calvetti_and_somersalo_2008, calvetti_et_al_2020a}, also adopt an MRF structure, particularly effective for representing sharp features in the solution.

The modeling of Student's \stT distribution in terms of a GSM has been explored in the previous literature, with seminal works dating back to \cite{chu_1973, andrews_and_mallows_1974}. So-called hierarchical \stT formulations for MRFs are briefly discussed in \cite[Ch. 5]{rue_and_held_2005}. Furthermore, applications of Gibbs sampling based on the GSM formulation of Student's \stT distribution have been investigated across various domains, including macroeconomic time series \cite{geweke_1993}, random effects models \cite{choy_and_smith_1997}, and empirical data analysis \cite{fernandez_and_steel_1998}.

\paragraph{Contributions and structure of the paper}
The main highlights of this work are outlined as follows:
\begin{itemize}
	\item[(i)] We introduce a class of flexible priors based on MRF structure combined with Student's \stT distribution. These priors can accommodate the reconstruction of either sharp or smooth features in the solution depending on the available data. 
	\item[(ii)] We reach such flexibility by solving a hierarchical Bayesian inverse problem that also estimates the degrees of freedom parameter of  Student's \stT distribution, for which we analyze different types of informative priors. 
	\item[(iii)] We exploit the GSM representation of Student's \stT distribution to sample the resulting hierarchical posterior more efficiently using the Gibbs sampler. 
	\item[(iv)] We compare the accuracy and performance of the Gibbs sampler with the results obtained by Hamiltonian Monte Carlo based on its \emph{No-U-Turn Sampler} (NUTS) adaptation \cite{hoffman_and_gelman_2014}. The performance of our method is demonstrated through numerical experiments on linear inverse problems.
\end{itemize}

This article proceeds as follows. In \Cref{sec:fram}, we introduce the mathematical framework for finite-dimensional Bayesian inverse problems; we present an approach for encoding structural information based on MRFs and discuss GSM representation. In \Cref{sec:priors}, we define our priors by combining the structure of first-order MRFs with the family of Student's \stT distribution; we also discuss how GSM can be used to represent these distributions and facilitate the posterior inference. \Cref{sec:comp} presents the computational framework and introduces two algorithms used to sample from the posterior distribution: Gibbs and NUTS. Numerical examples illustrating the proposed method are described in \Cref{sec:numexp}. \Cref{sec:conclusions} provides the main findings of this study.

\section{Preliminaries}\label{sec:fram}
We begin with a mathematical background introducing a discrete perspective on Bayesian inverse problems and its formulation for encoding structural information using MRFs. We also include a succinct presentation on GSM models.

\subsection{Discrete Bayesian inverse problems}
We consider the inverse problem of finding unknown parameter functions of a mathematical model that are defined over a domain $\mathcal{D}$, using noisy observed data of an underlying physical process described by the mathematical model. This represents an infinite-dimensional inverse problem that we approach by first discretizing all functions and then using a finite-dimensional perspective. In this case, the dimension $d$ of the discretized function will depend on the discretization size $N$, for example, in one- and two-dimensional problems $d=N$ and $d=N^2$ (assuming equal discretization in both directions), respectively. 

We employ the Bayesian approach to inverse problems \cite{kaipio_and_somersalo_2005}. The unknown parameter function is represented as a discretized random field given by a random vector $\ve{X}$ taking values $\ve{x}\in\mathcal{X}:=\mathbb{R}^d$ indexed on $\mathcal{D}$. Hence, we assume the distribution of $\ve{X}$ is absolutely continuous with respect to the Lebesgue measure, and it has a well-defined \emph{prior} probability density $p_{}(\ve{x})$. 

Now let $\mathcal{Y}:=\mathbb{R}^m$ be the data space. The mathematical model is represented via a forward operator $\matr{A}:\mathcal{X}\to\mathcal{Y}$ coupling parameters and data. We assume to have data $\ve{y} \in \mathcal{Y}$ arising from observations of some $\ve{x}$ under $\matr{A}$ corrupted by additive observational noise $\ve{e}\in\mathcal{Y}$, that is
\begin{equation}
\ve{y} = \matr{A}(\ve{x}) + \ve{e},\qquad \ve{e} \sim \mathcal{N}(\ve{0},\matr{\Sigma}_{\mathrm{obs}}),
\end{equation}
where the noise covariance matrix is defined as $\matr{\Sigma}_{\mathrm{obs}}:= \sigma_\mathrm{obs}^2\matr{I}_m$ with noise variance $\sigma_\mathrm{obs}^2$ and identity matrix $\matr{I}_m$ of size $m$. Since the noise is assumed to be Gaussian, and further assuming that the noise and model parameters are independent, the \emph{likelihood} is defined as 
\begin{equation}\label{eq:like}
p_{}(\ve{y}\given\ve{x}) =\frac{1}{(2\pi )^{\nicefrac{m}{2}}\sigma_\mathrm{obs}^m} \exp\left(-\frac{1}{2\sigma_{\rm obs}^2}\norm{\ve{y}-\matr{A}(\ve{x})}_2^2\right).
\end{equation}

The solution of the Bayesian inverse problem is the \emph{posterior} distribution; Bayes' theorem provides a way to construct such posterior as
\begin{equation}\label{eq:Bayes}
p_{}\left(\ve{x}\given \ve{y}\right) = \frac{1}{Z}\, p_{}(\ve{y}\given\ve{x})\,p_{}(\ve{x}),
\end{equation}
where $Z = \int_{\mathcal{X}} p_{}(\ve{y}\given\ve{x})\,p_{}(\ve{x})\, \dd \ve{x}$ is the normalization constant.

Having defined the posterior, we are typically faced with the task of extracting information from it. This is in general not a trivial task in high dimensions. One practical option is to employ variational methods, which assume that most of the information resides in a single peak of the posterior, e.g., the \emph{maximum a posteriori} (MAP) estimator, which is a point $\ve{x}_{\scriptscriptstyle \rm MAP}$ that maximizes posterior \cref{eq:Bayes}. Depending on the forward operator and the choice of the prior, this task can be a non-convex optimization problem; we likely find local maximizers that can still be used to build approximations to the posterior. Another method for characterizing the posterior in high dimensions is sampling. The idea is to generate a set of points $\{\ve{x}_i\}_{i=1}^{n}$ distributed, sometimes only approximately, according to $p_{}\left(\ve{x}\given \ve{y}\right)$. This enables the estimation of several summary statistics via Monte Carlo methods, in particular the \emph{posterior mean} (PM), which is an optimal point estimate $\ve{x}_{\scriptscriptstyle \rm PM}$ in the sense that it minimizes the Bayes risk (see, e.g., \cite[p.88]{tenorio_2017}).

Our focus is to estimate posterior distributions arising in imaging science, including cases where the unknown solutions can have either sharp or smooth features. These characteristics can be modeled via prior distributions that have the flexibility to express different features in the unknown solution. 

\begin{remark}\label{rem:01}
The interplay between the noise variance and the (scale-squared) parameters of the prior essentially defines the so-called regularization parameter used in classical deterministic inverse problems to control the quality of the solution. In practice, however, the parameter $\sigma_{\rm obs}^2$ is usually unknown, and hence, we decide to model it as a random variable and infer it from the data. In this case, we use an inverse-gamma hyperprior $\sigma_{\rm obs}^2\sim \invg(a,b)$, with shape parameter $a=1$ and scale parameter $b=10^{-4}$, following the recommendation in \cite{higdon_2007}.
\end{remark}

\subsection{Encoding structural information}
In infinite-dimensional Bayesian inverse problems, priors are usually constructed directly in functional spaces. Following this approach, Gaussian priors promoting the smooth behavior can be represented using, e.g., the Karhunen--Lo{\`e}ve expansion \cite{spanos_and_ghanem_1989}. Besov space priors are defined on a wavelet basis and, unlike Gaussian priors, are able to produce discontinuous samples, which is useful in problems requiring edge preservation \cite{lassas_et_al_2009}. Neural networks can also be used to impose smooth or sharp behaviors depending on the distribution chosen for the weights and biases of the network \cite{li_et_al_2022}.

Since our focus lies in discrete Bayesian inverse problems, we follow a different approach in which the involved functions are first discretized, and then priors are constructed on those discretizations. Under this formulation, MRF models can be used to incorporate structural information into the prior. MRFs are defined in connection with an undirected graph that contains vertices and edges associating dependencies between elements of the solution vector. For example, when the distribution of the MRF is Gaussian, the graph is defined by the nonzero pattern of the precision matrix \cite[p.21]{rue_and_held_2005}. When there is a strong dependency between neighboring elements of the solution, a common way to define the precision matrix is via the forward difference matrix (with zero boundary condition)
\begin{equation}\label{eq:diff}
\matr{D}=\begin{bmatrix}
	1 &   &        &    &   \\
	-1 &  1 &        &    &   \\
	& -1 &      1 &    &   \\
	&    & \ddots & \ddots &   \\
	&    &        & -1 & 1 \\
\end{bmatrix}_{N\times N},
\end{equation}
where $N$ is the discretization size. In two-dimensional applications, we often distinguish between horizontal and vertical differences, $\matr{D}^{(1)} = \matr{I}_N\otimes \matr{D},\quad \matr{D}^{(2)} = {\matr{D}} \otimes \matr{I}_N$, respectively ($\otimes$ denotes Kronecker product and $\matr{I}_N$ is a square identity matrix of size $N$). We remark that in practice one works with \emph{intrinsic} Gaussian MRFs, which have precision matrices not of full rank. Moreover, since we utilize first-order forward differences, the intrinsic MRF is known to be of the first order as its precision matrix has rank $d-1$ \cite[p.87]{rue_and_held_2005}.

Using the pairwise difference structure set via Equation \cref{eq:diff}, we can define a new parameter vector $\ve{u}$ containing differences
\begin{equation}\label{eq:diffvec}
\ve{u}=\matr{L}\ve{x},\qquad \text{where}\quad
\begin{cases}
\matr{L}=\matr{D} & \text{in one-dimensional cases},\\
\matr{L}=\begin{bmatrix}
	\matr{D}^{(1)}\\
	\matr{D}^{(2)}
\end{bmatrix} & \text{in two-dimensional cases}.
\end{cases}
\end{equation}

Then, the posterior \cref{eq:Bayes} can be reformulated as one of the following densities
\begin{subequations}
\begin{align}
p\left(\ve{u}\given \ve{y}\right) &\propto p(\ve{y}\given\matr{L}^{-1}\ve{u},\sigma_{\rm obs}^2)\,p(\ve{u})\label{eq:Bayes2a},\\
p\left(\ve{x}\given \ve{y}\right) &\propto p(\ve{y}\given\ve{x},\sigma_{\rm obs}^2)\,p(\matr{L}\ve{x}).\label{eq:Bayes2b}
\end{align}
\end{subequations}

The posterior \cref{eq:Bayes2a} uses a reparameterization of the inverse problem in terms of the unknown parameter differences $\ve{u}$, which facilitates computations involving the prior term; however, it does complicate the likelihood term as one requires inverting the difference matrix $\matr{L}$, which is not straightforward in other than one-dimensional problems (see, e.g., \cite{calvetti_and_somersalo_2007}). The posterior \cref{eq:Bayes2b} keeps the unknown parameter $\ve{x}$ as an objective of inference; however, computations of the prior term are more involved (see, e.g.,\cite{bardsley_2019}). Both formulations are essentially targeting the same problem, however, we focus on the second parameterization, as it does not require special methods to invert matrix $\matr{L}$.

We have seen that MRF models can be used to set a prior that incorporates strong dependence information. We still require the specification of the probability distribution of the MRF. Since we aim at defining a flexible model that can also account for smooth features in the solution, we consider a special yet rather wide class of distributions --- the Student's \stT distribution. We intend to combine Student's \stT distribution with the MRF structure presented in this subsection. The main advantage is that such a distribution family can be controlled by the \emph{degrees of freedom} parameter defining how heavy-tailed the distribution is. Samples from heavy-tailed distributions have larger jumps, compared to those sampled from Gaussian. The main difficulty is that resulting prior models can complicate the characterization of high-dimensional posteriors. To circumvent this, we explore GSM representations, which are discussed next.

\subsection{Gaussian scale mixtures}
The posterior can be examined in more depth by exploiting certain properties of the assumed priors. This is the case when the prior distributions can be expressed as a \emph{Gaussian scale mixture}, also known as scale mixtures of normals \cite{teichroew_1957, chu_1973, andrews_and_mallows_1974}. The GSM formulation plays a major role in hierarchical modeling, as it allows the expression of more involved distributions in a hierarchical manner by taking an integral of a Gaussian density.

\begin{definition}[Gaussian scale mixture]\label{def:mix}
Let $Y\sim\mathcal{N}(0,1)$ be a standard Gaussian random variable, and $Z$ a positive random variable with density $p_{\scriptscriptstyle Z}(z)$. Moreover, $Y$ and $Z$ are independent. We say that the density of the random variable $X=Y\,Z^{\nicefrac{1}{2}}$ given by
\begin{equation}\label{eq:GSM}
p_{\scriptscriptstyle X}(x) = \int_{0}^{\infty} p_{\scriptscriptstyle X\vert Z}(x\given z) p_{\scriptscriptstyle Z}(z)\,\dd z \quad \text{with}\quad p_{\scriptscriptstyle X\vert Z}(x\given z)=\frac{1}{\sqrt{2\pi z}}\exp\left(-\frac{1}{2}\frac{x^2}{z}\right),
\end{equation}
is a GSM with mixing density $p_{\scriptscriptstyle Z}(z)$ and Gaussian component density $p_{\scriptscriptstyle X\vert Z}(x\given z)$.
\end{definition}

Necessary and sufficient conditions for the existence of a GSM representation are discussed in \cite{andrews_and_mallows_1974}. They essentially amount to some positivity condition on the derivatives of $p_{\scriptscriptstyle X}(x)$, and also $p_{\scriptscriptstyle X}(x)$ has to be both symmetric and unimodal.

Random variables written as GSMs are usually referred to as sub-Gaussian, i.e., their distributions have strong tail decay, at least as fast as the tails of a Gaussian \cite[p.77]{samorodnitsky_and_taqqu_1994}. For example, many important symmetric random variables can be constructed using a GSM: the logistic and Laplace distributions \cite{andrews_and_mallows_1974}, generalized Gaussian distributions \cite{west_1987}, as well as, the Student's \stT and $\alpha$-stable families of distributions \cite[p.173]{feller_1971}. We summarize some of these constructions in \Cref{tab:GSM_list}.

\begin{table}[!ht]
\centering
\small
\begin{tabular}{l|l|ll}
\hline
Distribution & Gaussian & Mixture &  Parameters \\
$p_{\scriptscriptstyle X}(x)$ & $p_{\scriptscriptstyle X\vert Z}(x\given z)$ & $p_{\scriptscriptstyle Z}(z)$ &  \\
\hline\hline
\multirow{2}{*}{Laplace: $\mathcal{L}\left(\mu, \scaleT\right)$} & \multirow{2}{*}{$\mathcal{N}(\mu, z)$} & \multirow{2}{*}{Exponential: $\mathcal{E}\left(\lambda\right)$} & rate: \\
& &  & $\lambda=1/(2\tau^2)$ \\
\hline
\multirow{2}{*}{Student's \stT: $\mathcal{T}_{\nu}\left(\mu, \scaleT\right)$} & \multirow{2}{*}{$\mathcal{N}(\mu, \scaleT^2 z)$} & \multirow{2}{*}{Inverse-gamma: $\invg\left(\alpha, \beta\right)$} & shape and rate: \\
& &  & $\alpha=\beta=\nu/2$ \\
\hline
\multirow{2}{*}{$\alpha$-stable: $\mathcal{S}_\alpha\left(0, \mu, \scaleT\right)$} & \multirow{2}{*}{$\mathcal{N}(\mu, 2\scaleT^2 z)$} & \multirow{2}{*}{$\alpha$-stable: $\mathcal{S}_{\alpha/2}\left(\beta, \widetilde{\mu}, \widetilde{\scaleT}\right)$} & skewness, location, scale: \\
& &  & $\beta=1,\widetilde{\mu}=0,\widetilde{\scaleT}=1$ \\
\hline
\end{tabular}
\caption{Important examples of GSMs (see, e.g., \cite{choy_and_smith_1997}).} % Here the distributions have parameters location and scale, while the Gaussian components have parameters mean and variance.}
\label{tab:GSM_list}
\end{table}

When GSMs are used as priors for Bayesian inference, they induce a hierarchical structure, which is conditionally Gaussian at the first layer, and the mixing random variable becomes a hyperparameter at the second layer (see, e.g., \cite{choy_and_smith_1997, wainwright_and_simoncelli_1999}). This representation is ideal in iterative algorithms such as \emph{Markov chain Monte Carlo} (MCMC), since upon the introduction of the auxiliary mixing variable, the target random variable becomes conditionally Gaussian. Standard and efficient techniques are available to deal with Gaussian distributions compared to those required to analyze the more involved sub-Gaussian distributions. 

\section{Student's \stT MRF prior based on Gaussian scale mixture representation}\label{sec:priors}
We introduce a prior that combines the MRF structure with the parametric family of Student's \stT distributions (see also \cite[Ch.5]{rue_and_held_2005}). This distribution is characterized by tail-heaviness, which is modulated through the degrees of freedom parameter. Subsequently, we outline a method to explore the posterior distribution under this prior, with a particular focus on the modeling of the degrees of freedom parameter.

\subsection{Formulation of the prior}
We investigate a parametric family of MRF priors designed to capture both sharp and smooth features. This family utilizes the Gosset distribution, commonly known as Student's \stT distribution or simply the \stT distribution (we use these names interchangeably throughout the presentation). We employ this general family as a prior for each component in the difference random vector
\begin{equation}\label{eq:models}
{u}_{i}= [\matr{L}\ve{x}]_i\sim \mathcal{T}_{\nu}\left(\mu, \scaleT^2\right), \qquad \text{for}~i=1,\ldots,d, 
\end{equation}
where $\mathcal{T}_{\nu}\left(\mu, \scaleT^2\right)$ denotes Student's \stT distribution, and it has density function \cite[p.49]{feller_1971}
\begin{equation}
p(u) = \frac{c(\nu)}{\sqrt{\scaleT^2}} \left(1+\frac{(u-\mu)^2}{\nu\scaleT^2}\right)^{-\frac{(\nu+1)}{2}},
\end{equation}
here the number of degrees of freedom $\nu\in\mathbb{R}_{>0}$ defines the rate at which the tails of the distribution decrease, $\mu\in\mathbb{R}$ is the location parameter, $\scaleT^2\in\mathbb{R}_{>0}$ is the scale-squared parameter, and $c(\nu)={\Gamma(\nicefrac{(\nu +1)}{2})}/{\left(\Gamma(\nicefrac{\nu}{2})\sqrt{\pi\nu}\right)}$ is a normalization constant (with $\Gamma$ denoting the gamma function). 

\begin{figure}[!ht]
    \centering
    \includegraphics[width=0.6\linewidth]{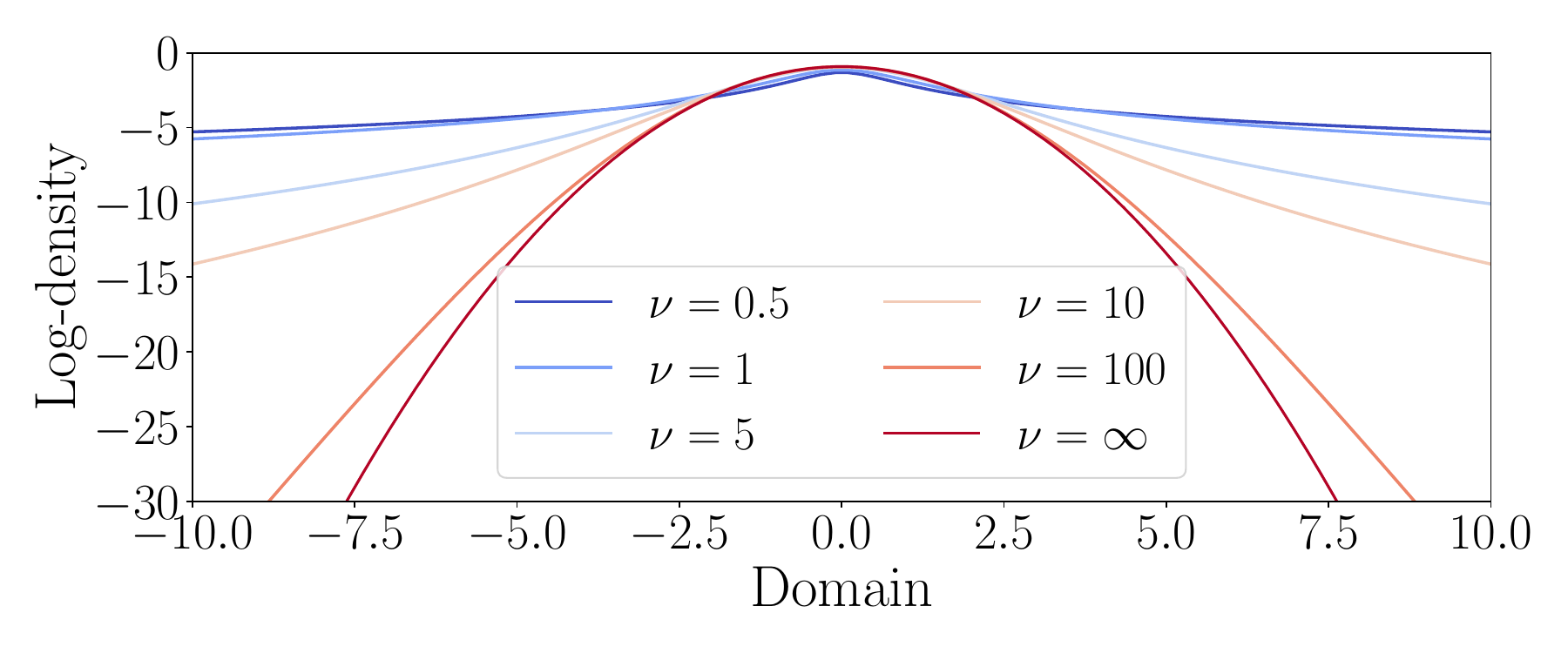}
    \caption{Logarithm of the probability density of the univariate standard Student's \stT distribution $\mathcal{T}_{\nu}\left(0, 1\right)$ for different values of the degrees of freedom parameter $\nu$. As $\nu \to \infty$, the \stT distribution $\mathcal{T}_{\infty}\left(0, 1\right)$ is equivalent to the univariate standard Gaussian distribution $\mathcal{N}(0,1)$.}
    \label{fig:student_t}
\end{figure}

Notably, in the limit $\nu\to \infty$, the \stT distribution approaches a Gaussian distribution. Conversely, when $\nu=1$, the distribution corresponds to a Cauchy distribution. It is worth noting that in practical applications, the degrees of freedom parameter is typically constrained to values greater than one or two, ensuring a well-defined mean or variance for the prior distribution, respectively. Consequently, to emphasize edge preservation, attention is directed towards distributions with lower degrees of freedom; in the opposite case, when aiming to represent smooth features, distributions with higher degrees of freedom become necessary. Intuitively, when using heavy-tailed densities we are increasing the probability of occurrence of large jump events (i.e., we are promoting large values of $\ve{u}$, thereby inducing sharpness). To illustrate the relationship between the degrees of freedom and the tail-heaviness of the \stT distribution, we visually depict in \Cref{fig:student_t} the logarithm of the density function of the univariate standard \stT distribution $\mathcal{T}_{\nu}\left(0, 1\right)$ with increasing degrees of freedom.

Sampling from posterior distributions in high dimensions under the Student's \stT MRF prior is potentially cumbersome. To improve the efficiency of the prior implementation, one can utilize the GSM representation of Student's \stT distribution. As we anticipated in \Cref{tab:GSM_list}, the mixing distribution is in this case inverse-gamma \cite{andrews_and_mallows_1974}. By incorporating this property, we can transform the original prior model into a hierarchical one, where the target difference random vector follows (conditionally) a more tractable Gaussian distribution. Therefore, we can write the GSM prior density for the $i$-th component of difference random vector as follows
\begin{subequations}\label{eq:GSM_t0}
\begin{align}
&p({u}_i) = \int_{0}^{\infty} p({u}_i\given \tau^2,{w}_i^2)p({w}_i^2) \,\dd{w}_i^2,\label{eq:GSM_t} \\
&({u}_i\given \tau^2,{w}_i^2) \sim \mathcal{N}\left({\mu},\tau^2w_i^2\right), \quad
{w}_i^2 \sim \invg\left(\nicefrac{\nu}{2},
\nicefrac{\nu}{2}\right),\label{eq:GSM_t2}
\end{align}
\end{subequations}
where ${w}_i^2$ is the mixing random variable and $\invg(a,b)$ denotes the inverse-gamma distribution with shape parameter $a$ and scale parameter $b$. The inverse-gamma probability density is defined as
\begin{equation}
p(w_i^2) = \frac{b^a}{\Gamma(a)}(w_i^2)^{-a-1}\exp\left(-\frac{b}{w_i^2}\right);
\end{equation}
from the GSM formulation \cref{eq:GSM_t2}, we see that the shape and rate parameters are $a=b=\nu/2$.

By utilizing the GSM in Equation \cref{eq:GSM_t2} and the linear relation described in Equation~\cref{eq:diffvec}, we can establish a conditionally Gaussian MRF prior for the random vector $\ve{X}$. In Gaussian MRFs, the precision matrix plays a crucial role, as its sparsity structure determines the neighborhood system that explains conditional relationships among the elements of $\ve{X}$. In edge-preserving inverse problems, the precision matrix is constructed based on the differences between the elements of the parameter vector. Specifically, it depends on the matrix $\ve{L}$ in Equation \cref{eq:diffvec} (for detailed information, refer to \cite[p.68]{bardsley_2019}). Consequently, we can express the prior precision matrix as
\begin{equation}\label{eq:t_prec}
\matr{\Lambda}(\tau^2,\ve{w}^2) = \matr{L}^\tran\matr{W}(\tau^2,\ve{w}^2)\matr{L}, \qquad \matr{W}(\tau^2,\ve{w}^2) := \diag\left(\frac{1}{\tau^2w_1^2}, \ldots,\frac{1}{\tau^2w_d^2}\right). 
\end{equation}

As a result, by using the GSM representation of the \stT distribution on the random vector of differences $\ve{U}$, we establish a hierarchical prior for all the unknowns as follows
\begin{subequations}
\begin{align}
& p(\ve{x},\scaleT^2,\ve{w}^2,\nu) = p(\ve{x}\given\scaleT^2, \ve{w}^2)p(\ve{w}^2\given \nu)p(\nu)p(\scaleT^2),\label{eq:t_prior} \\
& (\ve{x}\given \scaleT^2,\ve{w}^2) \sim \mathcal{N}\left(\ve{\mu}, \matr{\Lambda}^{-1}(\tau^2,\ve{w}^2)\right),  \\
& \scaleT^2 \sim \invg\left(a,b\right), \quad
{w}_i^2 \given \nu \sim \invg\left(\nicefrac{\nu}{2},\nicefrac{\nu}{2}\right), \quad \nu \sim p(\nu), \label{eq:t_priorcomp}
\end{align}
\end{subequations}
where the squaring $\ve{w}^2$ applies element-wise, and $p(\nu)$ denotes the prior on the degrees of freedom parameter $\nu$, which is discussed next.

\begin{remark}\label{rem:02}
In Equation \cref{eq:GSM_t2}, each mixing variable $w_i^2$ appears to control the variance \emph{locally} for each difference component $u_i$. Analogously, the scale-squared parameter $\scaleT^2$ controls the variance \emph{globally} across the domain. Since this parameter is relevant in the prior specification, we assume it is uncertain and endow it with an inverse-gamma hyperprior $\scaleT^2\sim \invg(a,b)$. Analogously to the discussion in \Cref{rem:01}, we set the shape and scale parameters as $a=1$ and $b=10^{-4}$.
\end{remark}
\begin{remark}\label{rem:03}
The global-local scheme associated with the hierarchical prior \cref{eq:t_prior} shares similarities with the horseshoe prior \cite{uribe_et_al_2023}, however, the hyperpriors for the local and global scale parameters are defined as half-Cauchy distributions for which the GSM representation has a different form. Moreover, the hierarchical prior \cref{eq:t_prior} is also related to the models discussed in \cite{calvetti_et_al_2020b, law_and_zankin_2022}, which employ a generalized inverse-gamma distribution as hyperprior.
\end{remark}

\subsection{Modeling the degrees of freedom} \label{sec:dofs}
Estimation of the degrees of freedom parameter $\nu$ of Student's \stT distribution has been actively studied \cite{geweke_1993,fernandez_and_steel_1998,fonseca_2008}. The choice of prior distribution is known to be challenging as this parameter is typically poorly identified from the data \cite{villa_2013}. 

We consider two popular prior choices \cite{juarez_2010, lee_2022}:
\begin{enumerate}
\item $\gam(\nu; \, a=2, b=0.1, \mu=0)$, the gamma distribution with the shape parameter $a \in \mathbb{R}_{>0}$, rate parameter $b \in \mathbb{R}_{>0}$, and location parameter $\mu \in \mathbb{R}_{>0}$. The gamma probability density is defined as
\begin{equation}\label{eq:p_nu_gamma_rate}
p(\nu)  = \frac{b^a}{\Gamma(a)} (\nu-\mu)^{a-1}\exp\left[ -b(\nu - \mu)\right], \qquad \nu \geq \mu.
\end{equation}
\item $\log\mathcal{N}\left( \nu; \, \mu=1, \sigma=1 \right)$, the log-normal distribution with parameters $\mu\in \mathbb{R}$ and $\sigma^2\in \mathbb{R}_{>0}$. The log-normal probability density is defined as
\begin{equation}\label{eq:p_nu_log_normal}
p(\nu) = \frac{1}{\nu \sqrt{2\pi \sigma^2}}\exp\left( -\frac{(\log \nu - \mu)^2}{2\sigma^2}\right).
\end{equation}
\end{enumerate}

In computational statistics, it is recommended to set a lower limit $\nu > 1$ on the gamma distribution. This ensures the first moment (mean) of the \stT distribution exists. Therefore, in addition to the previous prior choices, we compare the following thresholded gamma distributions:
\begin{enumerate}
\addtocounter{enumi}{2}
\item $\gam(\nu; \, a=2, b=0.1, \mu=1)$, a thresholded Gamma distribution which is equivalent to the prior choice $1$, but now excluding values $\nu \leq 1$. We remark that in this case $\mu \neq 0$, and we further employ notation $\mathrm{Ga}_{\nu > \mu}(\nu; \, a, b)$ to emphasize the application of a threshold.
\item $\gam(\nu; \, a=3, b=0.1, \mu=1)$, a thresholded Gamma distribution with shape $a = 3$. This distribution has a mode at $20$ and puts more probability mass away from the interval $(1, 10)$ as compared to the first three prior options. Thus, it promotes larger values of $\nu$ resulting in smooth behavior rather than piecewise constant. This can be beneficial if the task is to reconstruct smooth features in the object rather than sharp edges.
\end{enumerate}

We illustrate the performance of the four chosen priors in \Cref{sec:numexp_deconv}.

\subsection{Posterior under the Student's \stT prior} \label{sec:post}
The Bayesian inverse problem of estimating the posterior \cref{eq:Bayes2b} under the hierarchical prior in Equation \cref{eq:t_prior} is written as the hierarchical Bayesian inverse problem of determining the posterior density
\begin{equation}\label{eq:t_Bayes_hrc}
p(\ve{x},\sigma_{\mathrm{obs}}^2,\scaleT^2,\ve{w}^2,\nu\given \ve{y}) = p(\ve{y}\given\ve{x}, \sigma_{\mathrm{obs}}^2) p(\sigma_{\mathrm{obs}}^2) \,p(\ve{x}\given\scaleT^2, \ve{w}^2)p(\scaleT^2)p(\ve{w}^2\given\nu)p(\nu).
\end{equation}

In particular, using the relation between joint, conditional, and marginal distributions \cite[p.157]{feller_1971}, one can verify that for each parameter, the full conditional densities associated with Equation \cref{eq:t_Bayes_hrc} are
\begin{subequations}\label{eq:conds}
\begin{align}
p\left(\ve{x}\given \sigma_{\mathrm{obs}}^2,\tau^2,\ve{w}^2\right) &\propto p(\ve{y}\given\ve{x},\sigma_{\mathrm{obs}}^2) p(\ve{x}\given\tau^2,\ve{w}^2),\label{eq:condt1}\\
p(\sigma_{\mathrm{obs}}^2\given \ve{x}) &\propto p(\ve{y}\given\ve{x},\sigma_{\mathrm{obs}}^2)p(\sigma_{\mathrm{obs}}^2),\label{eq:condt2}\\
p(\tau^2\given \ve{x}, \ve{w}^2) &\propto p(\ve{x}\given\tau^2,\ve{w}^2)p({\tau}^2),\label{eq:condt3}\\
p(\ve{w}^2\given \ve{x}, \tau^2,\nu) &\propto p(\ve{x}\given\tau^2,\ve{w}^2)p(\ve{w}^2\given \nu),\label{eq:condt4}\\
p(\nu\given\ve{w}^2) &\propto p(\ve{w}^2\given\nu)p(\nu).\label{eq:condt5}
\end{align}
\end{subequations}

Now we can replace the corresponding densities on each conditional to derive them in a closed form when possible. 

\emph{Conditional 1: $p\left(\ve{x}\given \sigma_{\mathrm{obs}}^2,\tau^2,\ve{w}^2\right)$.} This density can be seen as a linear Gaussian Bayesian inverse problem which has a closed-form solution for the posterior (see, e.g., \cite[p.78]{kaipio_and_somersalo_2005}). Therefore, we obtain 
\begin{equation}\label{eq:t_cond1}
\begin{split}
p\left(\ve{x}\given \sigma_{\mathrm{obs}}^2,\scaleT^2,\ve{w}^2\right) &\propto \exp\left(-\frac{1}{2} \left(\frac{1}{\sigma_{\mathrm{obs}}^2}\norm{\ve{y}-\matr{A}\ve{x}}_2^2 + \bigl\| \ve{\Lambda}^{\nicefrac{1}{2}}(\scaleT^2,\ve{w}^2)\,\ve{x} \bigr\|_2^2\right) \right)\\
\ve{x}\given \sigma_{\mathrm{obs}}^2,\scaleT^2,\ve{w}^2 & \sim \mathcal{N}\left(\widetilde{\ve{\mu}}(\scaleT^2,\ve{w}^2)\,,\, \widetilde{\ve{\Lambda}}^{-1}(\scaleT^2,\ve{w}^2)\right),
\end{split}
\end{equation}
with the precision and mean parameters
\begin{equation}\label{eq:t_cond1_params}
\widetilde{\ve{\Lambda}}(\scaleT^2,\ve{w}^2) = \frac{1}{\sigma_{\mathrm{obs}}^2} \matr{A}^\tran\matr{A} + \matr{\Lambda}(\scaleT^2,\ve{w}^2),\qquad \widetilde{\ve{\mu}}= \widetilde{\ve{\Lambda}}^{-1}(\scaleT^2,\ve{w}^2)\left(\frac{1}{\sigma_{\mathrm{obs}}^2}\matr{A}^\tran\ve{y}\right);
\end{equation}
recall the prior precision matrix $\matr{\Lambda}(\scaleT^2,\ve{w}^2)$ is defined in Equation \cref{eq:t_prec}. 
The Cholesky factorization provides the most direct sampling algorithm for a Gaussian distribution. In this approach, a sample is obtained as $\ve{x}^\star= \widetilde{\ve{\mu}} + \widetilde{\matr{\Lambda}}^{-\nicefrac{1}{2}}\ve{u}$, where $\ve{u}\sim\mathcal{N}(\ve{0},\matr{I}_d)$ represents a standard Gaussian random vector, and $\widetilde{\matr{\Lambda}}^{\nicefrac{1}{2}}$ denotes a lower triangular matrix with positive real diagonal entries. However, the Cholesky strategy becomes computationally prohibitive due to the need for factorizing the matrix $\widetilde{\matr{\Lambda}}(\tau^2,\ve{w}^2)$. This difficulty also arises in MCMC algorithms for Gaussian distributions, such as preconditioned Crank--Nicolson \cite{cotter_et_al_2013}. To overcome this, we can use Krylov subspace methods to sample from the Gaussian conditional distribution \cref{eq:t_cond1}. In particular, we can draw a sample from $p\left(\ve{x}\given \sigma_{\mathrm{obs}}^2,\scaleT^2,\ve{w}^2\right)$ through randomized optimization. For further details on this formulation, refer to \cite[p.81]{bardsley_2019}. 

\emph{Conditional 2: $p(\sigma_{\mathrm{obs}}^2\given \ve{x})$.} The conditional density for the noise variance is proportional to the product of Gaussian and inverse-gamma densities. Through conjugate relations, such density is another inverse-gamma density that can be directly simulated
\begin{equation}
\begin{split}\label{eq:t_cond2}
p(\sigma_{\mathrm{obs}}^2\given \ve{x}) 
&\propto p(\ve{y}\given\ve{x},\sigma_{\mathrm{obs}}^2)p(\sigma_{\mathrm{obs}}^2)\\
&\propto (\sigma_{\mathrm{obs}}^2)^{-\nicefrac{m}{2}} \exp \left( -\frac{1}{2 \sigma_{\mathrm{obs}}^2} \norm{\ve{y}-\matr{A}\ve{x}}_2^2 \right) \left[ (\sigma_{\mathrm{obs}}^2)^{-a-1} \exp \left( -\frac{b}{\sigma_{\mathrm{obs}}^2}\right) \right] \\
&\propto (\sigma_{\mathrm{obs}}^2)^{-(\nicefrac{m}{2} + a)-1}\exp\left(- \left[\frac{1}{2}\norm{\ve{y}-\matr{A}\ve{x}}_2^2 + b \right] \frac{1}{\sigma_{\mathrm{obs}}^2} \right)\\
\sigma_{\mathrm{obs}}^2\given \ve{x} &\sim \invg\left(\frac{m}{2}+a, \frac{1}{2}\norm{\ve{y}-\matr{A}\ve{x}}_2^2 + b\right), 
\end{split}
\end{equation}
recall from \Cref{rem:01} that the shape and scale parameters of the noise variance hyperprior are defined as $a=1$ and $b=10^{-4}$, respectively.

\emph{Conditional 3: $p(\scaleT^2\given \ve{x}, \ve{w}^2)$.} The conditional density for the scale-squared parameter $\scaleT^2$ is another inverse-gamma density
\begin{equation}
\begin{split}\label{eq:cond_tau}
p(\scaleT^2\given \ve{x}, \ve{w}^2) 
&\propto p(\ve{x}\given\tau^2,\ve{w}^2)p({\tau}^2)\\
&\propto (\scaleT^2)^{-\nicefrac{k}{2}} \exp \left( -\sum_{i=1}^{k}\frac{([\matr{L}\ve{x}]_i-\mu)^2}{2\scaleT^2w_i^2} \right) \left[  (\scaleT^2)^{-a-1} \exp \left( -\frac{b}{\scaleT^2}\right) \right]\\
&\propto (\scaleT^2)^{-(\nicefrac{k}{2} + a) - 1} \exp\left(-\left[ \sum_{i=1}^{k}\frac{([\matr{L}\ve{x}]_i-\mu)^2}{2w_i^2} + b \right] \frac{1}{\scaleT^2}\right)\\
\scaleT^2\given \ve{x}, \ve{w}^2&\sim \invg\left(\frac{k}{2}+a, \sum_{i=1}^{d}\frac{([\matr{L}\ve{x}]_i-\mu)^2}{2w_i^2}+b\right), 
\end{split}
\end{equation}
where $k=\{d, 2d\}$ in one- and two-dimensional problems, respectively. Recall that the hyperprior shape and scale parameters are the same as those of the noise variance (cf. \Cref{rem:02}).

\emph{Conditional 4: $p(w_i^2\given \ve{x}, \scaleT^2,\nu)$.} Since the auxiliary parameters are independent, we can derive the conditional density for each component of $\ve{w}^2$ as follows
\begin{equation}
\begin{split}\label{eq:cond_w}
p(w_i^2\given \ve{x}, \scaleT^2,\nu) 
&\propto p(\ve{x}\given\tau^2,\ve{w}^2)p(\ve{w}^2\given \nu)\\
&\propto (w_i^2)^{-\nicefrac{1}{2}} \exp \left( -\frac{([\matr{L}\ve{x}]_i-\mu)^2}{2\scaleT^2w_i^2} \right) \left[ (w_i^2)^{-\nicefrac{\nu}{2}-1} \exp \left(- \frac{\nu}{2 w_i^2} \right)\right]\\
&\propto (w_i^2)^{-(\nicefrac{1}{2} + \nicefrac{\nu}{2}) - 1} \exp\left(-\left[ \frac{([\matr{L}\ve{x}]_i-\mu)^2}{2\scaleT^2} +\frac{\nu}{2}\right] \frac{1}{w_i^2}\right) \\
w_i^2\given \ve{x}, \scaleT^2,\nu &\sim\invg\left(\frac{\nu+1}{2},\, \frac{([\matr{L}\ve{x}]_i-\mu)^2}{2\scaleT^2}+\frac{\nu}{2}\right). 
\end{split}
\end{equation}

\emph{Conditional 5: $p(\nu\given\ve{w}^2)$.} Finally, the conditional density for the degrees of freedom parameter $\nu$ is written as
\begin{equation}\label{eq:cond_nu}
p(\nu\given\ve{w}^2)
\propto p(\nu)p(\ve{w}^2\given\nu)= p(\nu)\prod_{i=1}^{d} \frac{(\nicefrac{\nu}{2})^{\nicefrac{\nu}{2}}}{\Gamma(\nicefrac{\nu}{2})}({w}_i^2)^{-\nicefrac{\nu}{2}-1} \exp\left( -\frac{\nu}{2{w}_i^2}\right). 
\end{equation}

This one-dimensional density can be sampled within the Gibbs iterations using classical algorithms such as rejection sampling or \emph{random walk Metropolis} (RWM). In our study, after comparing both methods, we choose to use RWM. For this RWM-within-Gibbs sampling step, we can also perform a few extra within-Gibbs iterations (i.e., a within-Gibbs burn-in $\bar{n}_b$) in order to accelerate the convergence of the Gibbs chain and to reduce the correlation in the resulting posterior samples (see, e.g., \cite[p.213]{gamerman_and_lopes_2006}). To improve the sampling process, we use vanishing adaptation during the within-Gibbs burn-in \cite{andrieu_2008}. This type of adaptation ensures that the proposal is less and less dependent on the recent states in the chain, i.e., the adaptation gradually disappears from the proposal.

\section{Computational procedure for posterior sampling} \label{sec:comp}
We describe two MCMC algorithms to sample from the posterior distribution under the Student's \stT prior. The first algorithm, the Gibbs sampler, exploits the GSM representation of Student's \stT distribution. The other algorithm, NUTS, uses the traditional Student's \stT formulation; we consider NUTS to validate the numerical results obtained using Gibbs sampling. 

We remark that, from a practical perspective, we generate $n_s$ samples from the Markov chain, which are selected after discarding $n_b$ samples during the warm-up phase of the algorithm (so-called burn-in period). Moreover, to reduce storage requirements and improve the sample correlation, the sample chain values can be stored every $n_t$ iterations (lagging steps). Therefore, we need to draw $n=n_b+n_sn_t$ Markov chain steps to obtain $n_s$ samples from the posterior. 

\subsection{Gibbs sampler}
Gibbs sampling operates as an iterative algorithm that systematically selects a single variable and resamples it based on its conditional distribution, given the other variables. The sampling process is based on the fact that, under mild conditions, the set of full conditional distributions determines the joint distribution \cite{besag_1974}. Moreover, convergence conditions of the Gibbs sampler are discussed in \cite{roberts_and_smith_1994, tierney_1994}. By deriving the full conditional densities of all parameters, as shown in \Cref{sec:post}, we can use the Gibbs sampler to generate a Markov chain with a stationary distribution equivalent to the posterior distribution \cref{eq:t_Bayes_hrc}.

In Gibbs sampling, the process of choosing which parameter to sample (i.e., the variable index), often referred to as the \emph{scan order}, plays a crucial role \cite{owen_2023}. Two predominant scan orders are employed: random scan and systematic scan. In random scan, variables are chosen uniformly and independently at random during each iteration. In systematic scan, a predetermined permutation is chosen, with variables consistently selected in that specified order throughout the sampling process. Without loss of generality, we follow a random scan strategy; this is mainly due to the fact that this Gibbs sampler version produces a Markov chain that is reversible with respect to the posterior. The resulting Gibbs sampler applied to posterior \cref{eq:t_Bayes_hrc} is summarized (with a slight abuse of notation) in \Cref{alg:gibbs}. 

\RestyleAlgo{ruled}
\SetKwComment{Comment}{\# }{}
\begin{algorithm}[ht!]
\small
\caption{Random scan Gibbs sampler}\label{alg:gibbs}
\KwData{Gibbs parameters: number of samples $n_s$, burn-in $n_b$, thinning $n_t$. RWM-within-Gibbs parameters: warm-up $\bar{n}_b$.\\ \hspace{1.1cm} Posterior conditional densities \cref{eq:conds}. }
%\KwResult{Chains of parameters $\ve{x}$, $\sigma_\mathrm{obs}^2$, $\scaleT^2$, $\ve{w}^2$, $\nu$}
Initialize  $\ve{x}$, $\sigma_\mathrm{obs}^2$, $\scaleT^2$, $\ve{w}^2$, $\nu$ \;
$n \gets n_sn_t + n_b$\; %\Comment*[r]{total number of states to generate} 
$j \gets 1$ \;
\For{$i\leftarrow 1$ \KwTo $n$}{
  $k \leftarrow \texttt{randint}([1, 5])$ \Comment*[r]{generate the variable index}
  \Switch{$k$}{
    \uCase{$1$}{
        compute posterior mean $\widetilde{\ve{\mu}}(\scaleT^2,\ve{w}^2)$ and precision matrix $\widetilde{\ve{\Lambda}} \;(\scaleT^2,\ve{w}^2)$\;% according to \cref{eq:t_cond1_params} \;
        sample $\ve{x}\sim \mathcal{N}(\widetilde{\ve{\mu}}, \widetilde{\ve{\Lambda}}^{-1})$\; %$\ve{x} \sim p(\cdot \given \sigma_{\mathrm{obs}}^2,\tau^2,\ve{w}^2)$ \;%using the Cholesky factor of $\widetilde{\ve{\Lambda}}$ \; %as described in \cref{eq:chol_fact} \;
    }
    \lCase{$2$}{sample $\sigma_\mathrm{obs}^2 \sim p(\cdot \given \ve{x})$ in a closed form according to \cref{eq:t_cond2}}
    \lCase{$3$}{sample $\scaleT^2 \sim p(\cdot\given \ve{x}, \ve{w}^2)$ in a closed form according to \cref{eq:cond_tau}}
    \lCase{$4$}{sample $\ve{w}^2 \sim p(\cdot \given \ve{x}, \scaleT^2,\nu) $ in a closed form according to \cref{eq:cond_w}}
    \lCase{$5$}{sample $\nu \sim p(\cdot \given\ve{w}^2)$ in \cref{eq:cond_nu} using RWM with warm-up $\bar{n}_b$}
  }
  \If{$i > n_b$ {\bf and} $\mathrm{mod}(i, n_t) = 0$}{ 
  $\ve{x}^{(j)} \gets \ve{x}$, $(\sigma_\mathrm{obs}^2)^{(j)} \gets \sigma_\mathrm{obs}^2$, $(\scaleT^2)^{(j)} \gets \scaleT^2$, $(\ve{w}^2)^{(j)} \gets \ve{w}^2$, $\nu^{(j)} \gets \nu$ \;%\Comment*[r]{save current samples} 
  $j \gets j + 1$\;
  }
  
}
\KwRet{$\{\ve{x}^{(j)}\}_{j=1}^{n_s}$, $\{(\sigma_\mathrm{obs}^2)^{(j)}\}_{j=1}^{n_s}$}, $\{(\scaleT^2)^{(j)}\}_{j=1}^{n_s}$, $\{(\ve{w}^2)^{(j)}\}_{j=1}^{n_s}$, $\{\nu^{(j)}\}_{j=1}^{n_s}$
\end{algorithm}

\subsection{NUTS algorithm} \label{sec:nuts}
We introduced the Bayesian inverse problem of estimating the posterior \cref{eq:t_Bayes_hrc} under the MRF prior, based on the GSM representation of  Student's \stT distribution. The GSM formulation lends itself well to the application of the Gibbs sampler, due to the availability of conditional densities for each parameter.

Alternatively, we consider imposing Student's \stT distribution on the vector of increments $\ve{u} = \matr{L}\ve{x}$ without the GSM representation. In this case, the posterior \cref{eq:Bayes2b} can be modeled hierarchically as follows 
\begin{subequations}\label{eq:t_nuts_hrc}
\begin{align}
& p(\ve{x},\scaleT^2, \nu\given \ve{y}) = p(\ve{y}\given\ve{x})\,p(\matr{L}\ve{x}\given\scaleT^2, \nu)p(\scaleT^2)p(\nu),\label{eq:nuts_t_1} \\
& (\matr{L}\ve{x}\given \scaleT^2,\nu) \sim \mathcal{T}_{\nu}\left(\mu, \scaleT^2\right), \quad  \scaleT^2 \sim \invg(a, b), \quad \nu \sim p(\nu);
\end{align}
\end{subequations}
to sample from this posterior, we can employ Hamiltonian Monte Carlo based on its No-U-Turn Sampler adaptation \cite{hoffman_and_gelman_2014}. This method requires the logarithm of the posterior and also its gradient. 

Our formulation relies on linear forward models and Gaussian likelihoods, hence, the log-likelihood and its derivative with respect to the target parameter $[\ve{x},\scaleT^2,\nu]$ are well-known (see, e.g., \cite[p.11]{petersen_and_pedersen_2012}). However, as the prior distribution is more involved, we now write its logarithm and associated derivatives. The log-prior on $\ve{x}$ is
\begin{equation}\nonumber
\begin{split}
\log p(\matr{L}\ve{x}\given\scaleT^2, \nu) \propto & \, d \left[\log\Gamma\left(\frac{\nu+1}{2}\right) - \log\Gamma\left(\frac{\nu}{2}\right) - \frac{1}{2} \log(\pi \nu ) - \frac{1}{2} \log(\scaleT^2) \right] - \\
& \left( \frac{\nu+1}{2}\right) \sum_{i=1}^d \log \left( 1 + \frac{([\matr{L}\ve{x}]_i - \mu)^2}{\nu \scaleT^2}\right),
\end{split}
\end{equation}
and its partial derivatives with respect to the parameters $\ve{x}$, $\tau^2$ and $\nu$ are 
\begin{subequations} \nonumber
\begin{align}
    \frac{\partial}{\partial \ve{x}}  \log p(\matr{L}\ve{x}\given\scaleT^2, \nu) \propto & \, - \left( \frac{\nu + 1 }{1 + \frac{(\matr{L}\ve{x} - \mu)^2}{\nu \scaleT^2}} \right) \left(\frac{\matr{L}\ve{x} - \mu}{\nu\scaleT^2} \right)\matr{L}, \\
    \frac{\partial}{\partial \scaleT^2}  \log p(\matr{L}\ve{x}\given\scaleT^2, \nu) \propto & \,  - \frac{d}{2\scaleT^2} + \sum_{i=1}^d \left[\left(\frac{\nu+1}{2}  \right) \left( \frac{1 }{1 + \frac{(\matr{L}\ve{x} - \mu)^2}{\nu \scaleT^2}}  \right)\left( \frac{([\matr{L}\ve{x}]_i - \mu)^2}{(\nu (\scaleT^2))^2} \right)\right],  \\
    \frac{\partial}{\partial \nu}  \log p(\matr{L}\ve{x}\given\scaleT^2, \nu)  \propto & \, \frac{d}{2}\Psi\left(\frac{\nu+1}{2}\right) - \frac{d}{2}\Psi\left(\frac{\nu}{2}\right) - \frac{d}{2\nu} - \frac{1}{2} \sum_{i=1}^d \log \left( 1 + \frac{([\matr{L}\ve{x}]_i - \mu)^2}{\nu \scaleT^2}\right) + \nonumber \\
    & \frac{\nu+1}{2}  \sum_{i=1}^d  \left( \frac{1 }{1 + \frac{(\matr{L}\ve{x} - \mu)^2}{\nu \scaleT^2}} \right) \left( \frac{([\matr{L}\ve{x}]_i - \mu)^2}{\nu^2 \scaleT^2} \right),
\end{align}    
\end{subequations}
where $\Psi(\cdot)$ denotes the digamma function, or the logarithmic derivative of the gamma function $\Gamma$.

Finally, assuming a gamma prior $\gam(\nu; \, a, b)$ on the degrees of freedom parameter, we can follow the same procedure for the hyperparameters. The logarithms and derivatives of the hyperprior for $\scaleT^2$ and prior for $\nu$ are defined as
\begin{subequations}
\begin{align} \nonumber
\log p(\scaleT^2) & \propto (-a - 1) \log(\scaleT^2) - \frac{b}{\scaleT^2}, \quad 
&\frac{\partial}{\partial \scaleT^2} \log p(\scaleT^2) & \propto - \frac{a + 1}{\scaleT^2} + \frac{b}{(\scaleT^2)^2},\\
\log p(\nu) & \propto  (a - 1)\log(\nu) - b \,\nu, \quad &\frac{\partial}{\partial \nu} \log p(\nu) & \propto \frac{a-1}{\nu} - b. \nonumber
\end{align}
\end{subequations}

\section{Numerical experiments}\label{sec:numexp}
In the following, we illustrate the use of the proposed Student's \stT priors for Bayesian inversion. We consider linear inverse problems arising in signal processing and imaging science. The first example is a one-dimensional deconvolution problem that allows us to test multiple parameter settings in our computational framework. The second example is image deblurring. 

In our test problems, the point estimates $\bar{\ve{x}}$ (such as the posterior mean) of the target parameter are evaluated using the relative reconstruction error defined as
\begin{equation}
	\varepsilon_\mathrm{rel}:=\frac{\norm{\bar{\ve{x}}-\ve{x}^{\mathrm{true}}}_2}{ \norm{\ve{x}^{\mathrm{true}}}_2},
\end{equation}
where $\ve{x}^{\mathrm{true}}$ denotes the underlying true solution. Furthermore, the quality of the MCMC samples is measured by the \emph{effective sample size} (ESS). This metric reflects the amount of uncertainty within the chain due to autocorrelation. Therefore, if $\rho^{(i)}$ denotes the autocorrelation at the $i$-th lag, the ESS can be calculated as follows
\begin{equation}
    n_{\mathrm{eff}} = \frac{n_s}{1 + 2 \sum_{i=1}^{n_s}(\nicefrac{\rho^{(i)}}{\rho^{(0)}})} \approx \left\lceil \frac{n_s}{\tau_{\mathrm{int}}}\right\rceil, 
\end{equation}
where $\tau_{\mathrm{int}}$ stands for the integrated autocorrelation time (IACT). The definition of the ESS intuitively means that the larger the chain autocorrelation is, the smaller ESS is. Hence, while sampling one should target values of $n_{\text{eff}}$ as close as possible to the sample size $n_s$.

\subsection{One-dimensional deconvolution}\label{sec:numexp_deconv}
We consider the inverse problem of identifying a signal $x : [0, 1] \rightarrow \mathbb{R}$ from noisy convolved data. The mathematical model for convolution can be written as a Fredholm integral equation of the first kind
\begin{equation}\label{eq:fred_int_eq}
    y(t) = \int_0^1  \mathcal{A}(t, s) x(s) ds \quad \text{with} \quad \mathcal{A}(t, s) = \exp \left(-\frac{1}{2\sigma^2} (t - s)^2\right), \qquad 0 \leq t \leq 1,
\end{equation}
where $y(t)$ denotes the convolved signal and we employ a Gaussian convolution kernel $\mathcal{A}(t)$ with a fixed parameter $\sigma$ which controls the width of the blurring kernel.
% with parameter $s$ depending on the discretization size, i.e.  $ s = \round{N / 25} - 1$. 

A finite-dimensional representation of Equation~\cref{eq:fred_int_eq} is employed in practice. After discretizing the signal domain using $d = 130$ points, the convolution model can be expressed as a system of linear algebraic equations $\ve{y} = \matr{A}\ve{x}$. We consider synthetic observation data $\ve{y}$ with $m = d - 2$ equally-spaced elements (excluding two boundary points).

\begin{figure}[!ht]
    \centering
    \includegraphics[width=0.72\linewidth]{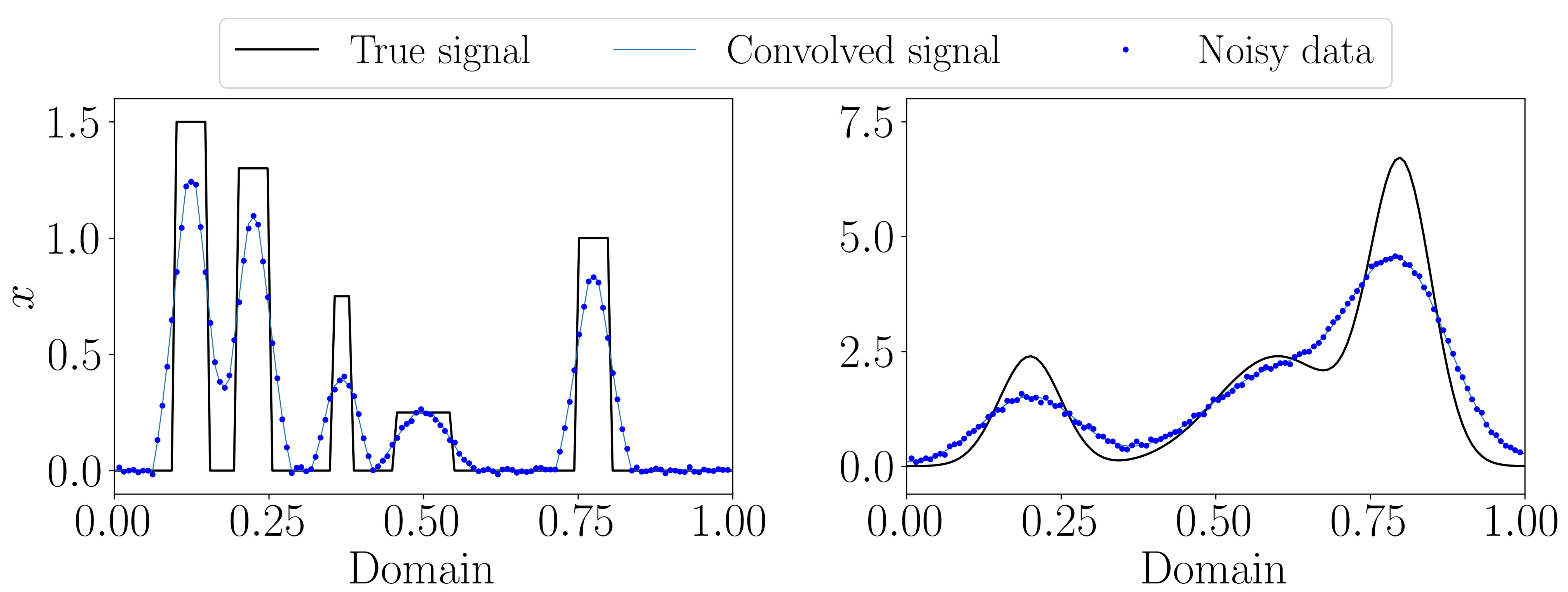}
    \caption{One-dimensional deconvolution example. Left: sharp signal; right: smooth signal.}
    \label{fig:deconv_1d}
\end{figure}

We compare two different signals, a piecewise constant (or sharp) signal and a smooth signal, to illustrate the flexibility of the proposed priors and their performance in recovering discontinuous signals as well as smooth. \Cref{fig:deconv_1d} shows the two underlying true signals together with the noisy observational data. The data were obtained according to model \cref{eq:fred_int_eq} with parameter $\sigma = 4$ for the piecewise constant signal and $\sigma = 8$ for smooth. The data were corrupted by adding Gaussian noise with standard deviation $\sigma_\mathrm{obs}^\mathrm{true} = 8.654 \times 10^{-3}$ for the piecewise constant signal and $\sigma_\mathrm{obs}^\mathrm{true} = 4.368\times 10^{-2}$ for the smooth signal.% (corresponding to $2\%$ relative error). 

First, we perform a prior study for the degrees of freedom parameter $\nu$ to select the most suitable option for our setting. Next, for the selected prior, we compare Gibbs and NUTS samplers and their efficiency in exploring the posterior distribution. Finally, we compare solutions obtained with the Student's \stT prior against the results obtained using two other popular priors --- the Laplace prior and the Cauchy difference prior. 

\emph{Prior comparison.} We study four different options for the prior distribution to model the degrees of freedom $\nu$, as described in \Cref{sec:dofs}. We compare two distribution options discussed in \cite{lee_2022}: gamma distribution $\gam(\nu; \, 2, 0.1)$ and log-normal distribution $\log\mathcal{N}\left( \nu; \, 1, 1 \right)$ with two thresholded variations of Gamma distribution: $\gamtr(\nu; \, 2, 0.1)$ and $\gamtr(\nu; \, 3, 0.1)$. 

We explore the resulting posterior distribution using the Gibbs sampler with the number of samples $n_s = 2 \times 10^4$, the global burn-in (or warm-up) period $n_b = 2 \times 10^3$, and lag (or thinning number) $n_t = 20$. To sample the degrees of freedom parameter $\nu$ within Gibbs, we use RWM sampler that produces one sample with burn-in period $\bar{n}_b = 100$; we choose this value based on numerical studies. Since the conditional of $\nu$ in \cref{eq:cond_nu} is computationally inexpensive to evaluate we can afford to perform extra within-Gibbs iterations (burn-in) to improve the resulting sub-chain of $\nu$.

\begin{figure}[!ht]
    \centering
    \includegraphics[width=0.82\linewidth]{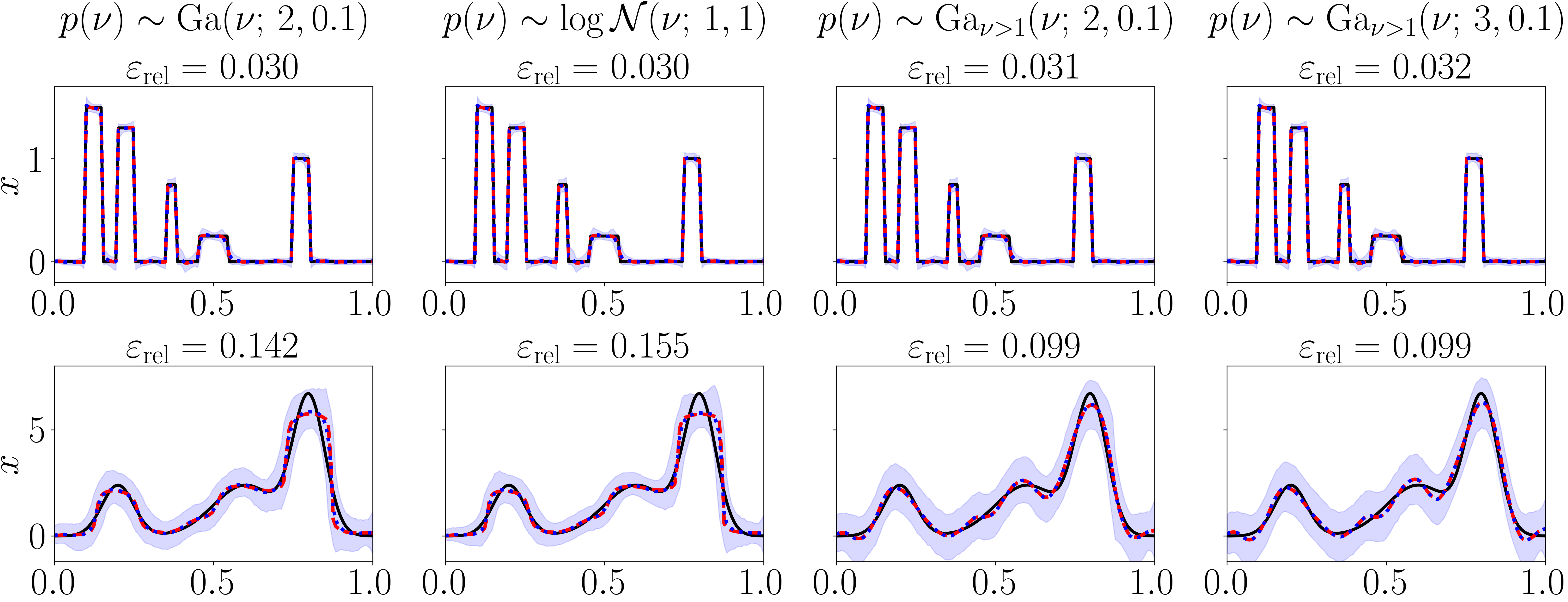}
    \caption{Comparison of posterior statistics for target parameter $\ve{x}$ using four different priors on the degrees of freedom $\nu$. Columns from left to right: 1)~gamma prior with shape $2$ and rate $0.1$; 2)~log-normal prior with scale and variance $1$; 3)~thresholded gamma prior with shape $2$ and rate $0.1$;  and 4)~thresholded gamma prior with shape $3$ and rate $0.1$. Top row: sharp signal, bottom row: smooth signal. True solution (black solid line), posterior median (red dashed line), posterior mean (blue dotted line), and $95\%$ highest density interval (shaded area).}
    \label{fig:hyperpr_test}
\end{figure}

\begin{figure}[!ht]
    \centering
    \includegraphics[width=0.82\linewidth]{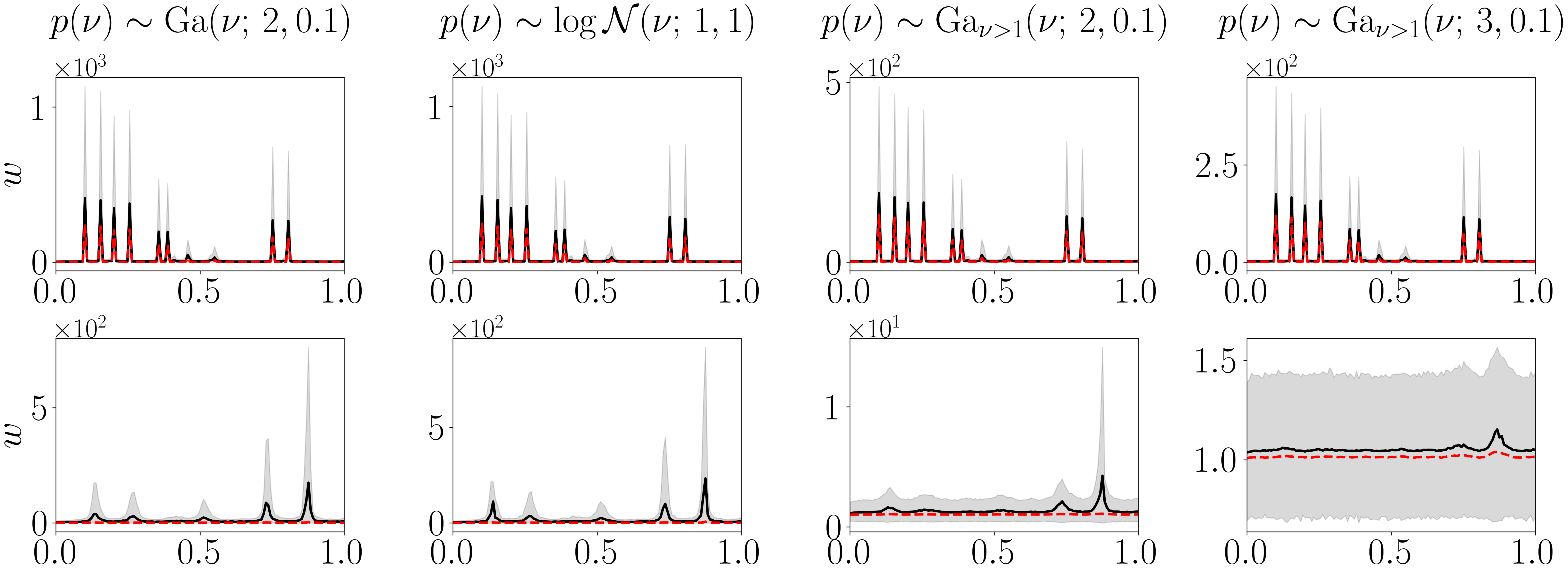}
    \caption{Comparison of posterior statistics for the local scale parameter $\ve{w}$ using four different priors on the degrees of freedom $\nu$. Columns from left to right: 1)~gamma prior with shape $2$ and rate $0.1$; 2)~log-normal prior with scale and variance $1$; 3)~thresholded gamma prior with shape $2$ and rate $0.1$;  and 4)~thresholded gamma prior with shape $3$ and rate $0.1$. Top row: sharp signal, bottom row: smooth signal. Posterior median (red dashed line) and mean (black solid line), and $95\%$ highest density interval (shaded area).}
    \label{fig:hyperpr_test_w}
\end{figure}

Posterior statistics for the target parameter $\ve{x}$ and the local scale parameter $\ve{w}$ for both piecewise constant and smooth signals are shown in~\Cref{fig:hyperpr_test,fig:hyperpr_test_w}, respectively. The piecewise constant signal is approximated with a smaller relative error, when the first two priors, $\gam(\nu; \, 2, 0.1)$ and  $\log\mathcal{N}\left( \nu; \, 1, 1 \right)$, are used since they allow for values $\nu < 1$ (which corresponds to more heavy-tailed behavior than Cauchy) as compared to the thresholded versions of Gamma distribution, $\gamtr(\nu; \, 2, 0.1)$ and $\gamtr(\nu; \, 3, 0.1)$, which restrict parameter $\nu$. In the smooth signal scenario, the first two priors result in worse signal reconstructions: note the poor approximation of the peak located at $0.8$. A possible explanation behind such artifacts in the solution is that we impose prior on the difference vector and, therefore, penalize changes in the curvature of the smooth signal. For the more suitable prior choices (priors $\gamtr(\nu; \, 2, 0.1)$ and $\gamtr(\nu; \, 3, 0.1)$), fewer artifacts are present in the smooth signal approximation and the reconstruction errors are smaller.

\begin{figure}[!ht]
    \centering
    \includegraphics[width=0.8\linewidth]{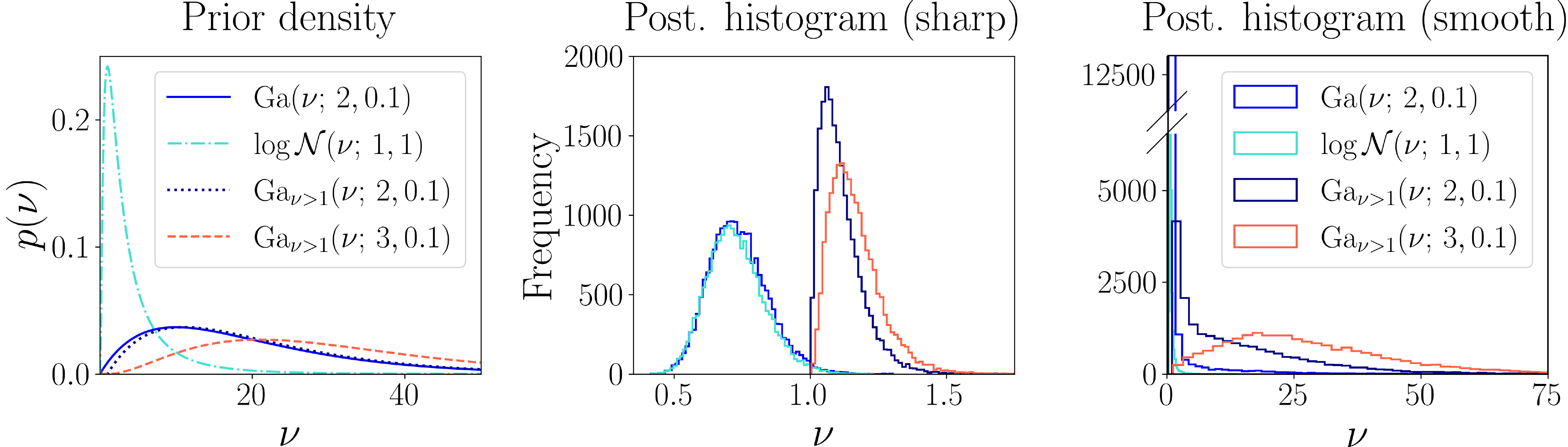}
    \caption{Prior and posterior of the degrees of freedom parameter $\nu$ (obtained using Gibbs). Left: different prior probability densities on $\nu$. Middle: posterior histograms of $\nu$ for the sharp signal. Right: posterior histograms of $\nu$ for the smooth signal.}
    \label{fig:gibbs_nu_pr_post}
\end{figure}

In \Cref{fig:gibbs_nu_pr_post}, we show the prior densities and corresponding histograms of posterior distributions for both signals. The first three prior distributions ($\gam(\nu; \, 2, 0.1)$, $\log\mathcal{N}\left( \nu; \, 1, 1 \right)$, and $\gamtr(\nu; \, 2, 0.1)$) have more probability mass placed on the interval $(1, 10)$ and, therefore, promote small values for the degrees of freedom parameter corresponding to the edge-promoting rather than smoothing behavior. On the other hand, distribution $\gamtr(\nu;\, 3, 0.1)$ has a mode at $20$ and assumes more mass far away from $1$. However, when prior $\gamtr(\nu;\, 3, 0.1)$ is used, we see that the posterior distribution for the smooth signal coincides with the prior distribution meaning that this prior is too strong and we do not explore posterior of $\nu$ at all. In general, the posterior histograms illustrate the flexibility of the prior to model different structural features. The model finds values of $\nu$ that better represent the data: in the piecewise constant case, it tends to favor small values of the degrees of freedom parameter, while in the smooth case, it tends to favor larger values. 

In \Cref{tab:gibbs_post_stats}, we report posterior statistics (sample mean, standard deviation, and effective sample size) for the scalar parameters: degrees of freedom $\nu$, global scale parameter $\tau$, and the noise standard deviation $\sigma_\mathrm{obs}$. Based on this analysis, we choose $\gamtr(\nu;\, 2, 0.1)$ as a prior distribution for the degrees of freedom parameter. This choice results in good reconstructions for both discontinuous and smooth cases and, in addition, allows for avoiding possible pathological behavior via restricting values of $\nu$. We remark that the values of the proposal scaling in the RWM-within-Gibbs after adaptation are around $0.555$ and $1.067$, for the sharp and smooth signals, respectively.

\begin{table}[!ht]
\centering
\small
\begin{tabular}{c|c|c|c|c|c}
\hline
Signal    & Prior for $\nu$                  & Parameter            & Mean                   & Std                    & $n_\mathrm{eff}$ \\ \hline\hline
          &                                  & $\nu$                 & $0.736$                & $0.107$                & $3\,954$         \\
          & $\gam(\nu; \, 2, 0.1)$           & $\scaleT$             & $6.269 \times 10^{-3}$ & $1.906 \times 10^{-3}$ & $1\,122$         \\
          &                                  & $\sigma_\mathrm{obs}$ & $8.397 \times 10^{-3}$ & $6.108 \times 10^{-4}$ & $12\,447$        \\ \cline{2-6} 
          &                                  & $\nu$                 & $0.726$                & $0.105$                & $4\,411$         \\
          & $\log\mathcal{N}(\nu; \, 1, 1)$  & $\scaleT$             & $6.179 \times 10^{-3}$ & $1.814 \times 10^{-3}$ & $1\,298$         \\
Sharp     &                                  & $\sigma_\mathrm{obs}$ & $8.401 \times 10^{-3}$ & $6.121 \times 10^{-4}$ & $14\,046$        \\ \cline{2-6} 
signal    &                                  & $\nu$                 & $1.112$                & $0.079$                & $4\,846$         \\
          & $\gamtr(\nu; \, 2, 0.1)$         & $\scaleT$             & $10.493 \times 10^{-3}$& $3.989 \times 10^{-3}$ & $684$            \\
          &                                  & $\sigma_\mathrm{obs}$ & $8.393 \times 10^{-3}$ & $6.185 \times 10^{-4}$ & $12\,683$        \\ \cline{2-6}
          &                                  & $\nu$                 & $1.165$                & $0.095$                & $3\,338$         \\
          & $\gamtr(\nu; \, 3, 0.1)$         & $\tau$                & $11.214 \times 10^{-3}$& $4.215 \times 10^{-3}$ & $622$            \\
          &                                  & $\sigma_\mathrm{obs}$ & $8.395 \times 10^{-3}$ & $6.185 \times 10^{-4}$ & $12\,803$        \\ \hline
          &                                  & $\nu$                 & $2.541$                & $5.987$                & $61$             \\
          & $\gam(\nu; \, 2, 0.1)$           & $\scaleT$             & $0.104$                & $0.167$                & $59$             \\
          &                                  & $\sigma_\mathrm{obs}$ & $4.365 \times 10^{-2}$ & $2.922 \times 10^{-3}$ & $17\,095$        \\ \cline{2-6}
          &                                  & $\nu$                 & $0.927$                & $0.605$                & $102$            \\
          & $\log\mathcal{N}(\nu; \, 1, 1)$  & $\scaleT$             & $0.050$                & $0.075$                & $81$             \\
Smooth    &                                  & $\sigma_\mathrm{obs}$ & $4.369 \times 10^{-2}$ & $2.900\times 10^{-3}$  & $15\,937$        \\\cline{2-6}  
signal    &                                  & $\nu$                 & $14.060$               & $13.679$               & $133$            \\
          & $\gamtr(\nu; \, 2, 0.1)$         & $\scaleT$             & $0.455$                & $0.186$                & $139$            \\
          &                                  & $\sigma_\mathrm{obs}$ & $4.341 \times 10^{-2}$ & $2.893 \times 10^{-3}$ & $17\,607$        \\ \cline{2-6}         
          &                                  & $\nu$                 & $28.198$               & $17.145$               & $464$            \\
          & $\gamtr(\nu; \, 3, 0.1)$         & $\scaleT$             & $0.573$                & $0.118$                & $925$            \\
          &                                  & $\sigma_\mathrm{obs}$ & $4.340 \times 10^{-2}$ & $2.880 \times 10^{-3}$ & $18\,920$        \\ \hline
\end{tabular}
\caption{Comparison of posterior statistics for scalar parameters $\nu$, $\tau$, and $\sigma_\mathrm{obs}$ using four different prior distributions on the degrees of freedom $\nu$. Mean, standard deviation (Std), and effective sample size ($n_\mathrm{eff}$).}\label{tab:gibbs_post_stats}
\end{table}

% SECOND STUDY
\emph{Sampler comparison: Gibbs vs. NUTS.} Using the prior distribution $\gamtr(\nu;\, 2, 0.1)$ on the degrees of freedom parameter $\nu$, we compare posterior statistics of $\nu$ obtained via the Gibbs sampler with those obtained via NUTS. In NUTS, we follow the standard formulation of the \stT distribution $\mathcal{T}_{\nu}\left(\mu, \scaleT^2\right)$ with hyperpriors $\scaleT^2 \sim \invg(1, 10^{-4})$ and $\nu \sim \gamtr(2, 0.1)$, as described in \Cref{sec:nuts}. We run NUTS to obtain $n_s = 2 \times 10^4$ states and apply additionally burn-in of $n_b = 2 \times 10^3$. 

\begin{figure}[!ht]
    \centering
    \subfloat{\includegraphics[width=0.85\linewidth]{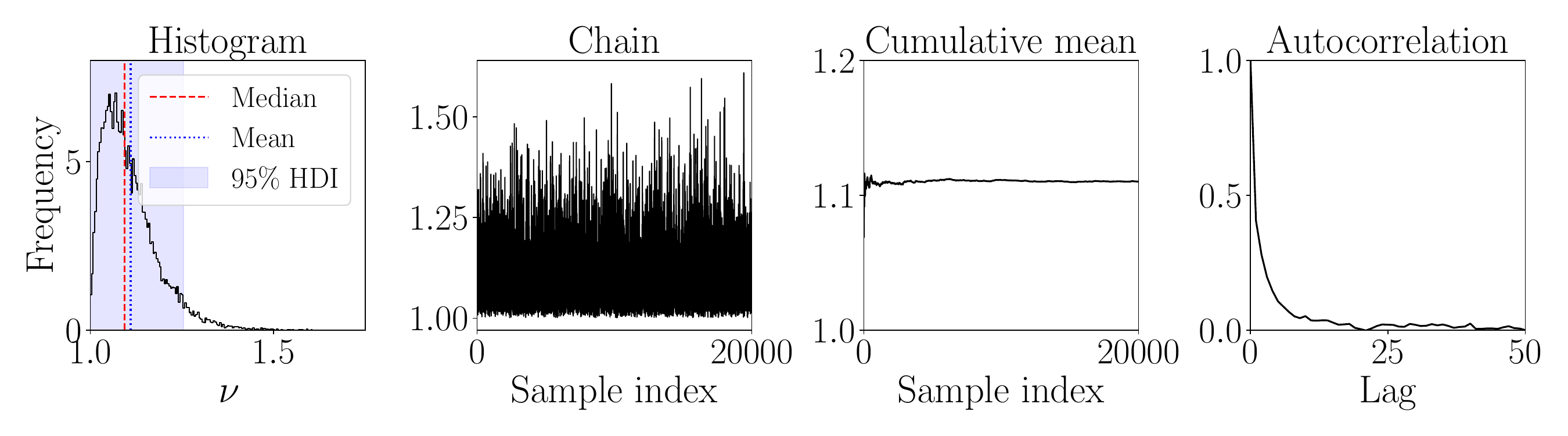}}\\
    \subfloat{\includegraphics[width=0.85\linewidth]{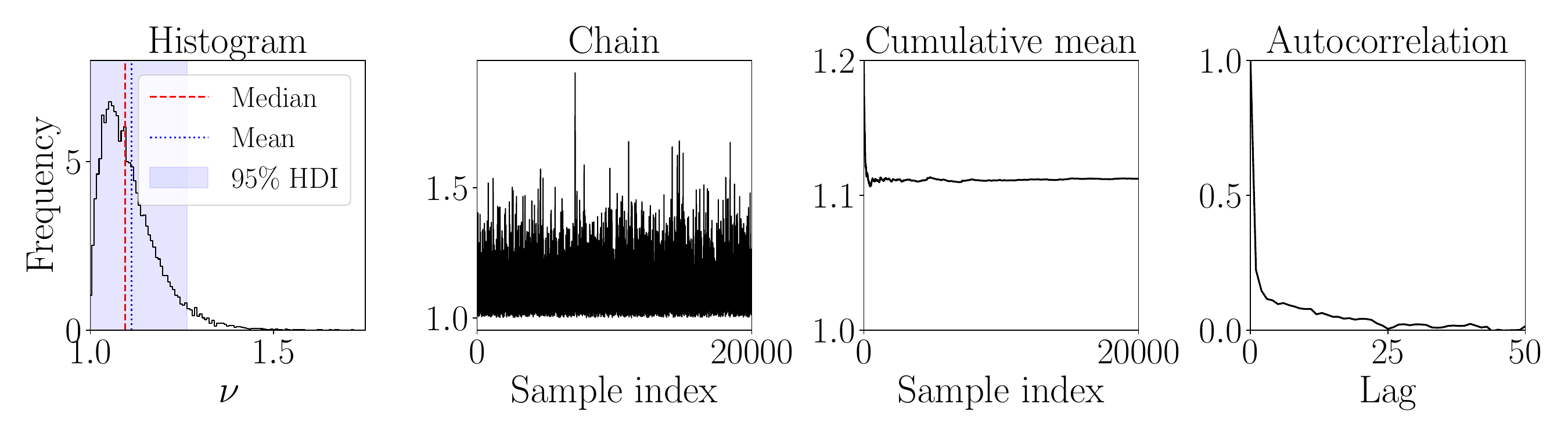}}
  \caption{Posterior statistics of $\nu$ for the \emph{sharp} signal obtained using NUTS (top) and Gibbs (bottom). Prior on $\nu$ is $\gamtr(2, 0.1)$.}
  \label{fig:nu_chains_datasky}
\end{figure}

Signal reconstructions are of the same good quality when either Gibbs or NUTS is used. In \Cref{fig:nu_chains_datasky}, we compare the posterior statistics for the parameter $\nu$ obtained using both samplers for the \emph{sharp} signal. Since the posterior chains are comparatively good, we can conclude that the Gibbs sampler is appropriately exploring the target posterior distribution. We omit the analogous chain plot for the smooth signal scenario for the sake of brevity. However, we note that inference of the $\nu$ parameter is more challenging in the smooth signal case, as the Gibbs chain has slightly more correlation than the one computed by NUTS. Nonetheless, we remark here that the main advantage of the Gibbs sampler over NUTS is its independence from gradient computations.

\emph{Prior comparison: Student's \stT prior vs. Laplace prior and Cauchy prior.} We also compare the posterior obtained using the Student's \stT prior with the posteriors obtained using the Laplace prior \cite{bardsley_2012} and the Cauchy difference prior \cite{markkanen_et_al_2019}. In the one-dimensional case, the Laplace and Cauchy difference priors are defined as 
\begin{equation}
p(\ve{x}) = \begin{cases}
    \left( \dfrac{1}{2\tau}\right)^d \exp\left(-\dfrac{\norm{\matr{D} \ve{x}}_1}{\tau}  \right) & \text{Laplace} \\
    \left( \dfrac{1}{\pi}\right)^d \displaystyle\prod\limits_{i = 1}^{d} \dfrac{\tau}{\left([\matr{D} \ve{x}]_i^2 + \tau^2\right)} & \text{Cauchy},
\end{cases} 
\end{equation}
where the difference matrix $\matr{D}$ is specified in \cref{eq:diff}, the parameter $\tau \in \mathbb{R}_{> 0}$ is the scale, and $\norm{\cdot}_1$ denotes the $\ell_1$-norm. Recall that the Cauchy difference prior is equivalent to the Student's \stT MRF prior with fixed $\nu = 1$. In this study, we sample both cases using NUTS ($n_s = 2 \times 10^4$ states with additional burn-in $n_b = 2 \times 10^3$). 

\begin{figure}[!ht]
    \centering
    \includegraphics[width=0.85\linewidth]{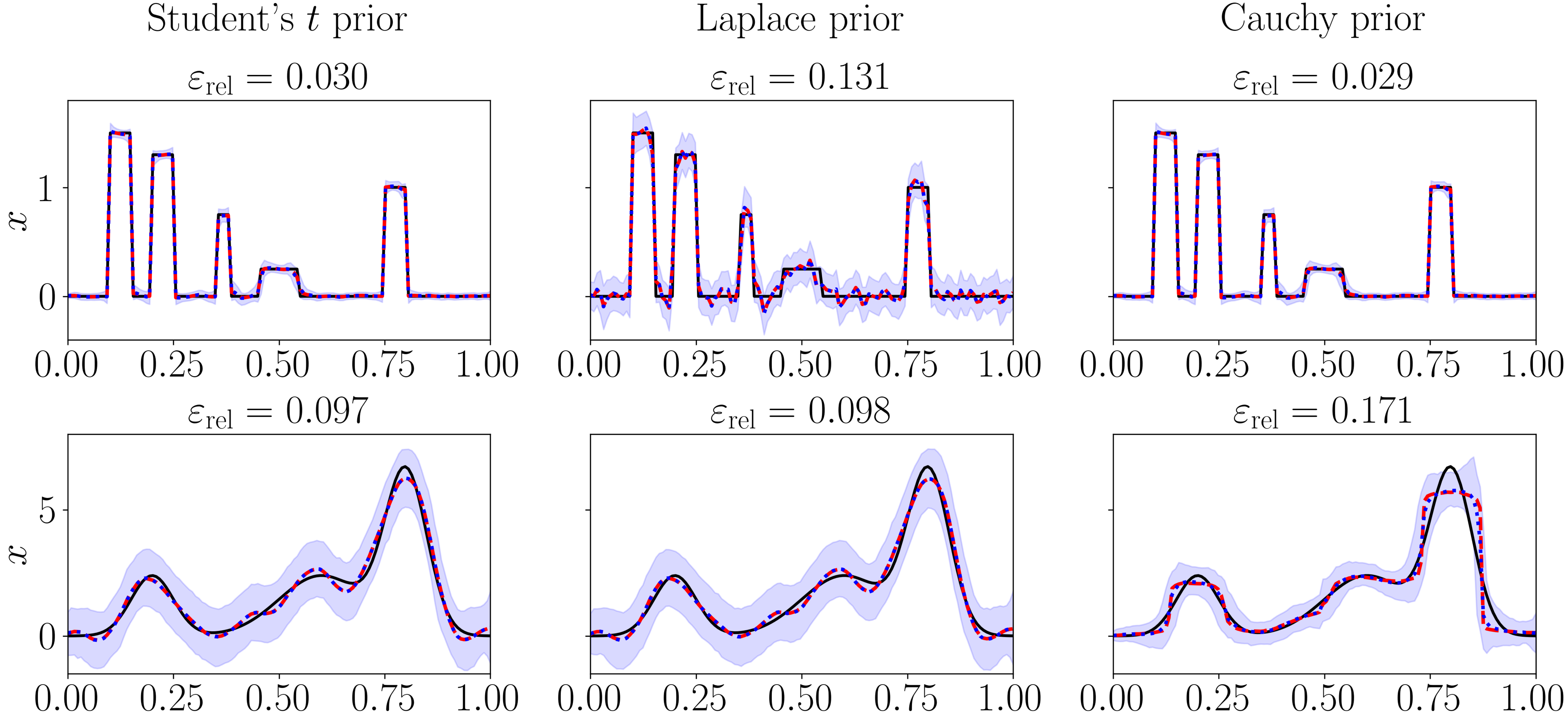}
    \caption{Comparison of posterior statistics for the target parameter $\ve{x}$ using NUTS. Left: Student's \stT prior, middle: Laplace prior, right: Cauchy prior. Results are shown for the sharp signal (top row) and smooth signal (bottom row). True solution (black solid line), posterior median (red dashed line), posterior mean (blue dotted line), and $95\%$ highest density interval (shaded area).}
    \label{fig:nuts_prior_test}
\end{figure}

In \Cref{fig:nuts_prior_test}, we show posterior statistics of parameter $\ve{x}$ for Student's $t$, Laplace, and Cauchy priors. In the piecewise constant case, the Student's \stT prior results in a similar reconstruction quality as compared to the Cauchy prior; whereas the \stT prior clearly reduces the reconstruction error and  the posterior uncertainty as compared to the Laplace prior. In the smooth signal case, the Student's \stT prior outperforms the Cauchy prior due to its flexibility, which allows the reconstruction of smooth features as well as sharp ones.

\begin{remark}
We also studied the influence of our Student's \stT MRF prior on the convergence of the sampling method under different discretization sizes in comparison to the Laplace prior. We used the sharp signal data set ($m=128$), considered sizes $d =\{64, 128, 256\}$, and estimated the posterior using NUTS ($n_s = 2 \times 10^4$ and $n_b = 2 \times 10^3$) for both priors. In each case, we computed five independent chains to check convergence using the estimated potential scale reduction factors (so-called $\widehat{R}$ diagnostic test) \cite{vehtari_gelman_2021}. For each discretization size and all the sample chains, we obtained $\widehat{R}$ values close to one for both priors. Thus, we assume that all our sample chains have converged. While the solutions may show some variation with different discretization sizes, this does not affect the convergence of the sampler.
\end{remark}

\subsection{Image deblurring}
In image deblurring, we aim to recover the original sharp image from the corresponding noisy blurred image using a mathematical model of the blurring process. If we denote the image domain by $\mathcal{D} = ([0,1]\times[0,1])$, and $\ve{t} = [t_1, t_2]^\tran \in \mathcal{D}$, $\ve{s} = [s_1, s_2]^\tran \in \mathcal{D}$ then the blurred image $y$ can be obtained from the true image $x$ using the Fredholm integral operator as follows \cite{lu_2010}
\begin{equation}\label{eq:fred_int_eq_2d}
    y(\ve{t}) = \int_{\mathcal{D}}  \mathcal{A}(\ve{t}, \ve{s}) x(\ve{s}) d\ve{s} \quad \mathrm{with} \quad \mathcal{A}(\ve{t}, \ve{s}) = \frac{1}{2\pi \sigma^2} \exp \left(-\frac{(t_1 - s_1)^2 + (t_2 - s_2)^2}{2\sigma^2} \right),
\end{equation}
where $\mathcal{A}(\ve{t}, \ve{s})$ is the Gaussian convolution kernel specifying the blurring (similar to the one-dimensional case).

\begin{figure}[!ht]
    \centering
    \includegraphics[width=0.6\linewidth]{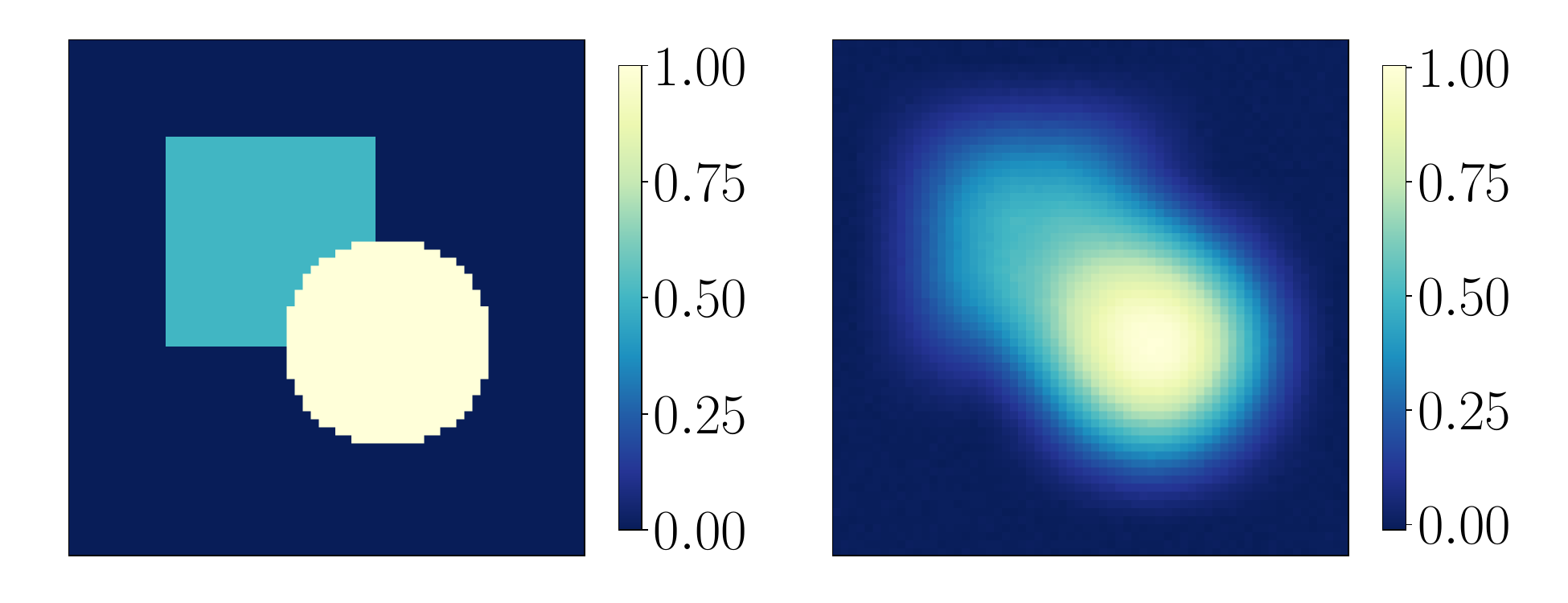}
    \caption{Image deblurring example. Left: underlying true image (square-and-disk phantom). Right: corresponding noisy blurred image (1\% relative noise).}
    \label{fig:deblur_problem}
\end{figure}

As a \emph{true} image, we use a square-and-disk phantom of size $64 \times 64$, that is, $N = 64$, and the problem dimension is $d = N^2 = 4\,096$ \cite{hansen_2010}. We generate a blurred image (of the same size) using the blurring model \cref{eq:fred_int_eq_2d} with parameter $\sigma = 6$ and standard deviation $\sigma_{\mathrm{obs}}^{\mathrm{true}} = 3.3 \times 10^{-3}$. The true image and its noisy blurry version are shown in \Cref{fig:deblur_problem}.

We estimate posterior under the Student's \stT prior (with $\nu \sim \gamtr(\nu;\, 2, 0.1)$) using the Gibbs sampler. Next, we compare results with those obtained by NUTS with Laplace MRF prior. This prior is defined in the two-dimensional case as 
\begin{equation}
    p(\ve{x}) = \left( \frac{1}{2\tau}\right)^d \exp\left[-\frac{1}{\tau}\left(\|\matr{D}^{(1)} \ve{x}\|_1 + \|\matr{D}^{(2)} \ve{x}\|_1\right) \right],
\end{equation}
where $\matr{D}^{(1)}$ and $\matr{D}^{(2)}$ are horizontal and vertical difference matrices, parameter $\tau \in \mathbb{R}_{> 0}$ is the scale, and $\norm{\cdot}_1$ denotes the $\ell_1$-norm. %We assume a value of the noise standard deviation $\sigma_{\mathrm{obs}}$ to be known and do not include it to the inference. 

\begin{figure}[!ht]
    \centering
    \includegraphics[width=0.6\linewidth]{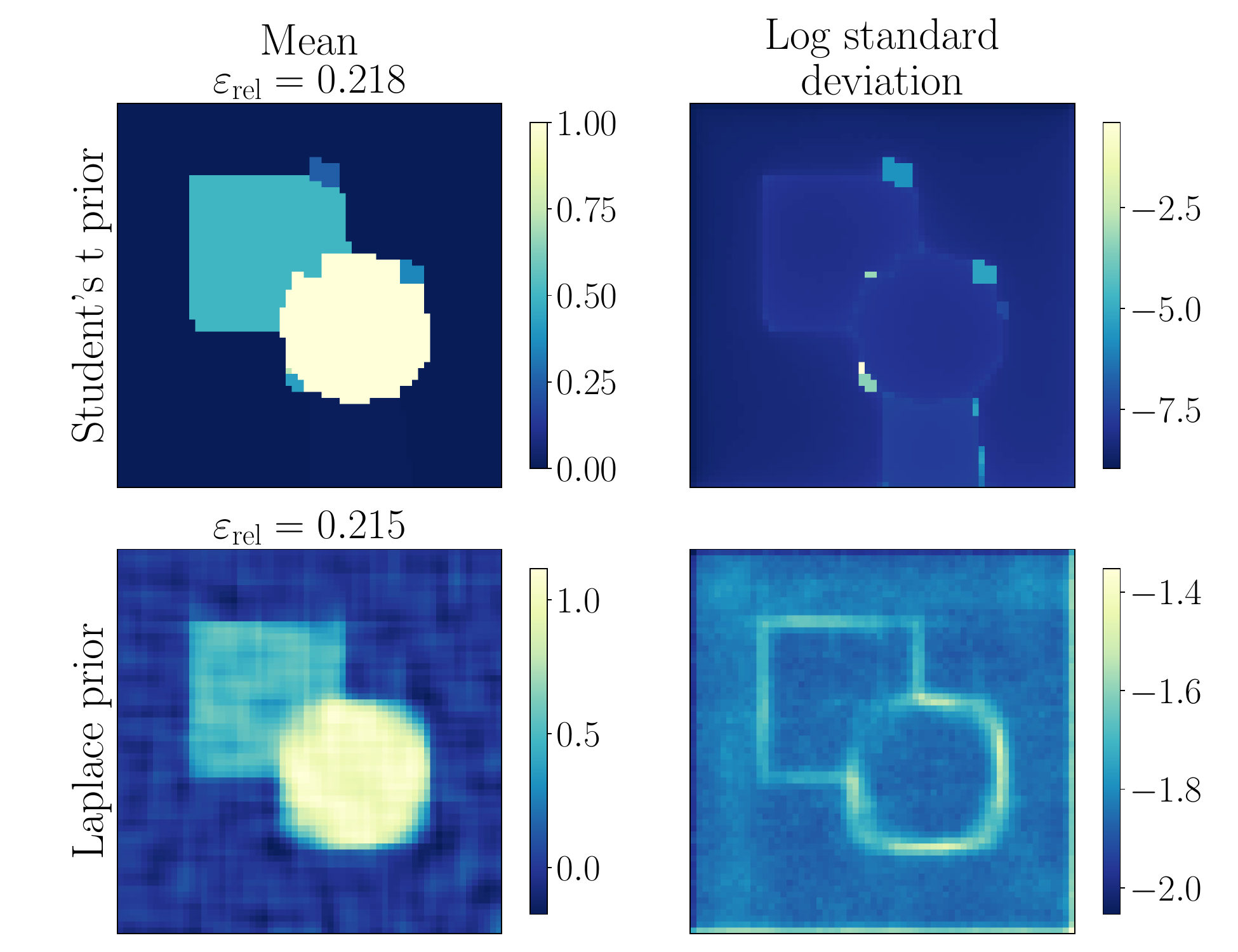}
    \caption{Comparison of posterior statistics for the target parameter $\ve{x}$ using the Student's \stT prior (top row) and the Laplace prior (bottom row). Posterior mean with relative error $\varepsilon_\mathrm{rel}$ (left) and the logarithm of the standard deviation (right).}
    \label{fig:deblur_x_param}
\end{figure}

The input parameters to the Gibbs and NUTS methods are the same as in the one-dimensional deconvolution example. In \Cref{fig:deblur_x_param}, we show the posterior mean and logarithm of standard deviation for two deblurred images obtained with Student's \stT prior and Laplace prior. We plot the \emph{logarithm} of standard deviation to facilitate the interpretation for the Student's \stT prior, as maximum values of standard deviation occur at single pixels of the image. We note for the Student's \stT prior, the minimum value of the standard deviation is $1.286 \times 10^{-4}$ and the maximum value is $0.041$ (if we omit two points exhibiting larger standard deviation values of $0.639$ and $0.675$); whereas for Laplace prior, the standard deviation ranges from $0.128$ to $0.259$. 

The relative reconstruction error is $0.218$ for the \stT prior and $0.215$ for the Laplace prior. Despite both priors yield comparable relative errors, we can clearly see that the mean of posterior based on the Student's \stT prior is sharper, contains less noise, and has smaller posterior uncertainty. It can be also noted that for the Laplace prior the larger values of standard deviation are located at the edges pointing to the higher uncertainty in the boundary regions.

\begin{figure}[!ht]
    \centering
    \includegraphics[width=0.6\linewidth]{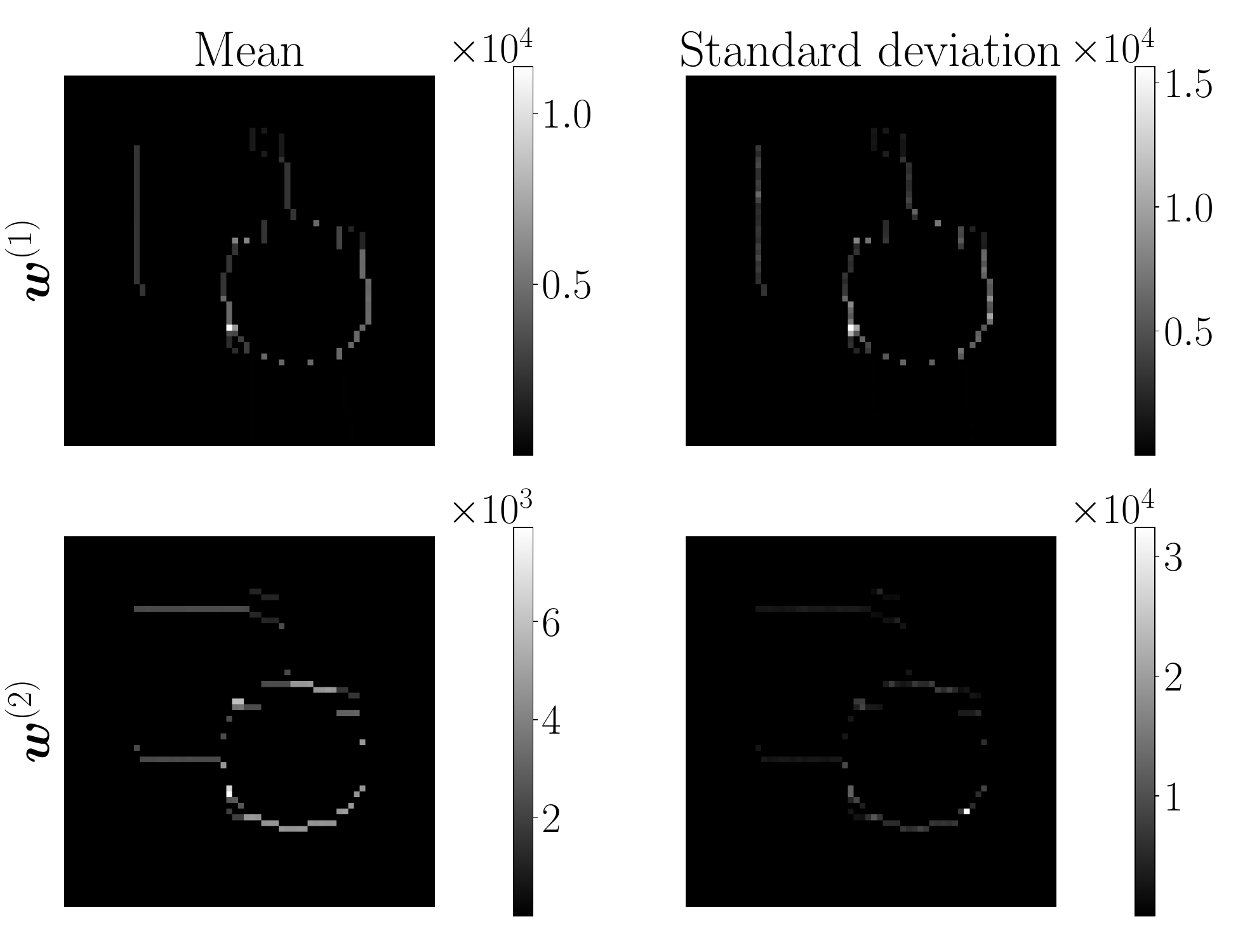}
    \caption{Posterior mean (left) and standard deviation (right) of the local scale parameter $\ve{w}$ obtained under the Student's \stT prior using Gibbs. Top row: horizontal components $\ve{w}^{(1)}$, bottom row: vertical components $\ve{w}^{(2)}$.}
    \label{fig:deblur_w_param}
\end{figure}

\begin{figure}[!ht]
    \centering
    \subfloat{\includegraphics[width=0.85\linewidth]{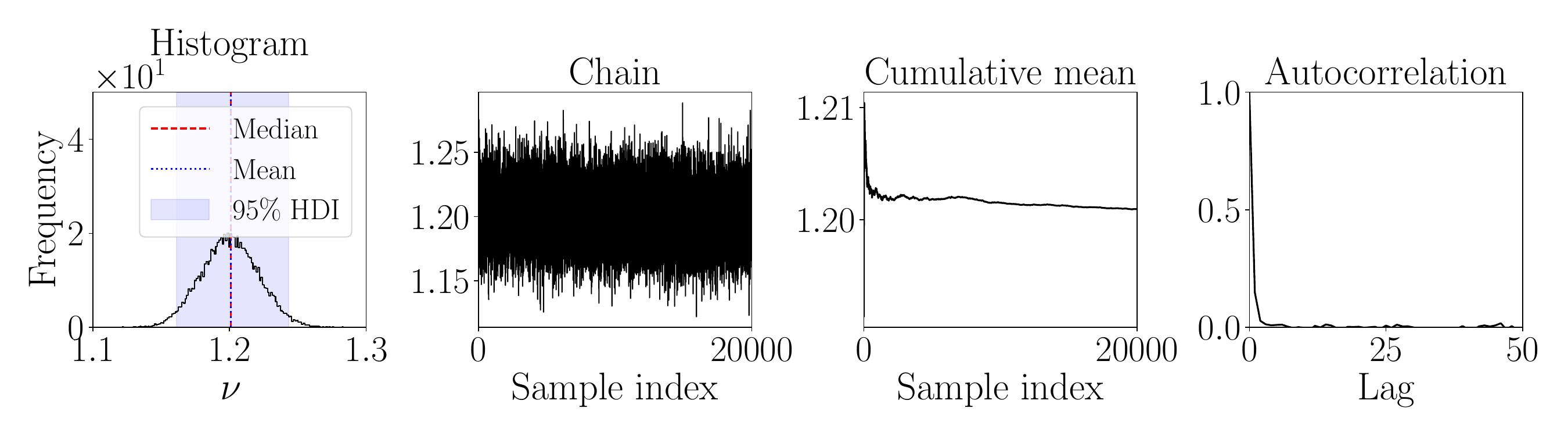}}\\
    \subfloat{\includegraphics[width=0.85\linewidth]{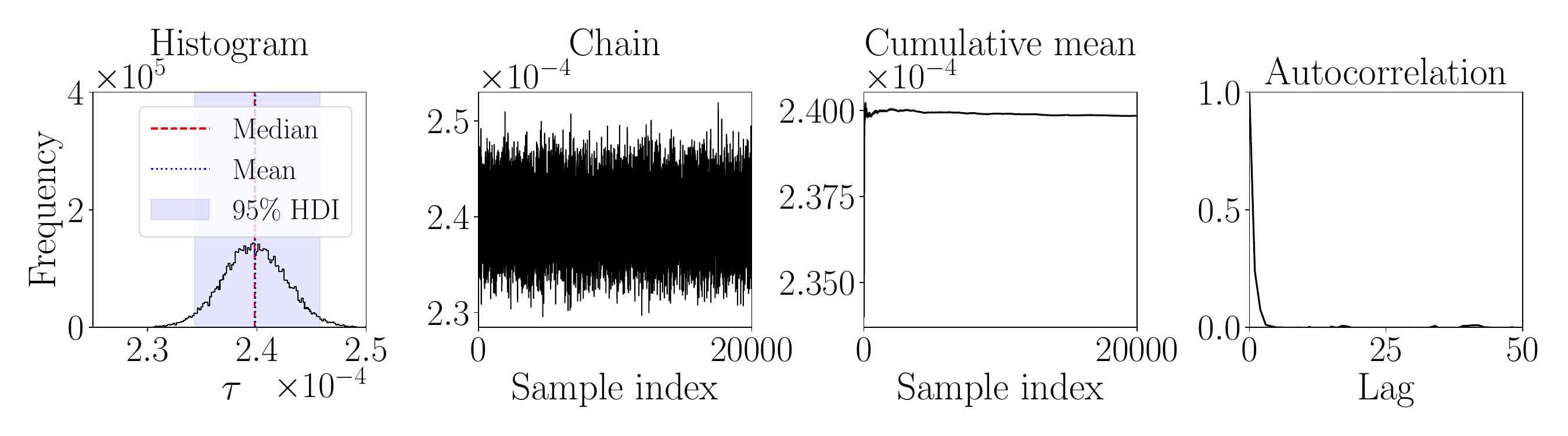}}
  \caption{Posterior statistics for $\nu$ (top row) and $\tau$ (bottom row), obtained under the Student's \stT prior using Gibbs: histogram, chain, cumulative mean, and sample autocorrelation.}
  \label{fig:deblur_post_chains}
\end{figure}

In addition, for the Student's \stT prior, we report posterior statistics of the local scale parameter $\ve{w}$ in \Cref{fig:deblur_w_param}, and chain statistics of scalar parameters (degrees of freedom $\nu$ and the global scale $\scaleT$) in \Cref{fig:deblur_post_chains}. In the Laplace prior case, the posterior mean and standard deviation of the scale parameter $\tau$ are $0.312$ and $0.015$, respectively. %we show posterior statistics of parameter $\delta$ (see \Cref{fig:deblur_laplace_delta}).

\section{Summary and conclusions}\label{sec:conclusions}
We developed a computational framework for solving linear Bayesian inverse problems, in which the unknown can be either discontinuous or smooth. Our method involves flexible priors based on Markov random fields combined with Student's \stT distribution. We solve a hierarchical Bayesian inverse problem that allows joint estimation of the degrees of freedom parameter of Student's \stT distribution. The estimation of the degrees of freedom tailors the prior for recovering not only sharp but also smooth features depending on the underlying unknown solution. Modeling the degrees of freedom as a hyperparameter, however, adds an extra layer to the hierarchical model, which increases the inference complexity. At the same time, this does not affect the computational complexity, as the parameter is one-dimensional. Moreover, we note that the automatic inference of the degrees of freedom parameter eliminates the need for expensive manual tuning of this parameter.

Due to the hierarchical structure and Student's \stT distribution involved in the prior, sampling of the posterior distribution is challenging. To address the issue, we employed a Gaussian scale mixture representation of Student's \stT distribution, that allowed us to employ an efficient Gibbs sampler. We validated numerically our suggested Gibbs sampler by comparing its results to the solution obtained via NUTS.

In the numerical experiments, we showed that Student's \stT hierarchical prior can be applied for recovering blocky objects as well as objects containing smooth features. To access the performance of the proposed method in Bayesian inversion, we compared Student's \stT priors with more traditional prior models based on Laplace and Cauchy Markov random fields commonly used for promoting sharp features. As compared to Laplace priors, Student's \stT priors enable the computation of sharp point estimates of the posterior while reducing the posterior uncertainty. At the same time, the \stT priors are flexible enough to capture smooth features in the solution, whereas the Cauchy MRF priors can only promote sharp behavior. We also observed that when the unknown signals are sharp, our model benefits from weak prior information (i.e., no truncation of the degrees of freedom hyperprior), while for smooth targets the prior information is preferably strong (i.e., truncation of the degrees of freedom hyperprior).

As a future work, we consider employing a higher-order difference matrix in the MRF structure to add more structural dependencies. This could be potentially beneficial for the smooth case scenario as opposed to our current implementation based on the first-order difference matrix, which models connections in the two neighboring pixels only. In addition, we would like to extend the method to the higher dimensional inverse problems. This essentially requires the application of efficient methods to sample from high-dimensional Gaussian distributions.

\section*{Acknowledgments}
This work was supported by the Research Council of Finland through the Flagship of Advanced Mathematics for Sensing, Imaging and Modelling, and the Centre of Excellence of Inverse Modelling and Imaging (decision numbers 359183 and 353095, respectively). In addition, AS was supported by the doctoral studies research grant from the \emph{Vilho, Yrj\"{o} and Kalle V\"{a}is\"{a}l\"{a} Foundation} of the Finnish Academy of Science and Letters.

\addcontentsline{toc}{section}{References}
\bibliographystyle{model1-num-names}
\bibliography{manuscript.bib}

\end{document}